\documentclass[structabstract,usenatbib]{aa}
\usepackage{txfonts}
\usepackage{graphicx}
\usepackage{url}
\usepackage{natbib}
\usepackage{rotating}
\bibpunct{(}{)}{;}{a}{}{,}	

\begin{document}

\title{The present-day mass function of the Quintuplet cluster\thanks{Based on observations collected at the ESO/VLT under Program ID 71.C-0344(A) (PI: F.~Eisenhauer, retrieved from the ESO archive) and Program ID 081.D-0572(B) (PI: W.~Brandner).}
\author{
B. ~Hu{\ss}mann\inst{\ref{inst1}}
\and A. ~Stolte\inst{\ref{inst1}}
\and W. ~Brandner\inst{\ref{inst2}}
\and M. ~Gennaro\inst{\ref{inst2}}
\and A. ~Liermann\inst{\ref{inst3}}
}
\institute{
	   Argelander Institut f\"ur Astronomie, Universit\"at Bonn, Auf dem H\"ugel 71, 53121 Bonn, Germany \email{[hussmann;astolte]@astro.uni-bonn.de}\label{inst1}
	   \and Max-Planck-Institut f\"ur Astronomie, K\"onigsstuhl 17, 69117 Heidelberg, Germany \email{[brandner;gennaro]@mpia.de}\label{inst2}
	   \and Max-Planck-Institut f\"ur Radioastronomie, Auf dem H\"ugel 69, 53121 Bonn, Germany \email{liermann@mpifr-bonn.mpg.de}\label{inst3}
	   }
\date{Received / Accepted}
\abstract{The stellar mass function is a probe for a potential dependence of star formation on the environment. Only a few young clusters are known to reside within the Central Molecular Zone and can serve as testbeds for star formation under the extreme conditions in this region.}{We determine the present-day mass function of the Quintuplet cluster, a young massive cluster in the vicinity of the Galactic centre.}{We use two epochs of high resolution near infrared imaging data obtained with NAOS/CONICA at the ESO VLT to measure the individual proper motions of stars in the Quintuplet cluster in the cluster reference frame. An unbiased sample of cluster members within a radius of $0.5\,\mathrm{pc}$ from the cluster centre was established based on their common motion with respect to the field and a subsequent colour-cut. Initial stellar masses were inferred from four isochrones covering ages from 3 to 5 Myr and two sets of stellar evolution models. For each isochrone the present-day mass function of stars was determined for the full sample of main sequence cluster members using an equal number binning scheme.}{We find the slope of the present-day mass function in the central part of the Quintuplet cluster to be  $\alpha = -1.66\pm0.14$ for an approximate mass range from $5$ to $40\,\mathrm{M_{\sun}}$, which is significantly flatter than the Salpeter slope of $\alpha = -2.35$. The flattening of the present-day mass function may be caused by rapid dynamical evolution of the cluster in the strong Galactic centre tidal field. The derived mass function slope is compared to the values found in other young massive clusters in the Galaxy.}{}
\keywords{Galaxy: center -- Galaxy: open clusters and associations: individual: Quintuplet cluster -- stars: luminosity function, mass function -- stars: early-type -- infrared: stars -- instrumentation: adaptive optics}}
\maketitle

\section{Introduction}
\label{s_introduction}

The Quintuplet cluster is one of only three young, massive clusters known within the central molecular zone (CMZ) with projected distances of less than $30\,\mathrm{pc}$ to Sagittarius~$\mathrm{A^*}$, the supermassive black hole (SMBH) at the centre of the Milky Way. The other two clusters are the Arches cluster at a similar location as the Quintuplet cluster, and the nuclear star cluster surrounding Sgr~$\mathrm{A^*}$. These clusters are unique laboratories to study the formation and evolution of stars and their host clusters in the Galactic Centre (GC) environment.

The conditions for star formation in the CMZ and the GC region are rather extreme in terms of high gas densities, enhanced temperatures, tidal forces exerted by the gravitational potential in the inner Galaxy, and strong magnetic fields.  These conditions were suggested to favour the formation of high mass stars as compared to the more moderate spiral arm environments \citep{Morris1993,Morris1996}. An overpopulation of high mass stars may be evidenced in a flattened \emph{initial} mass function (IMF) in GC star clusters. The young massive clusters are ideal candidates to search for such a deviation from the Galactic field IMF. Their youth ensures that a large fraction of the initial population is still present in or near the cluster and their high total masses provide coverage of the entire mass function (MF) up to the highest-mass stars known, such as the Pistol star in the Quintuplet cluster \citep{Figer1995,Figer1998,Yungelson2008}. Due to the large number of high mass stars these cluster are also well-suited to assess stellar evolution scenarios for the most massive stars.

A direct comparison of the observed present-day mass function (PDMF) of the GC young massive clusters with the Galactic field IMF is aggravated due to their rapid dynamical evolution and dissolution in the GC tidal field. N-body simulations of compact massive clusters with masses $\leq 2 \times 10^4 \,\mathrm{M_{\sun}}$ and distances from the GC $\leq 100\,\mathrm{pc}$ by \citet{Kim2000} yielded dissolution times of less than $10\,\mathrm{Myr}$. A similar study by \citet{Zwart2002} derived somewhat longer dissolution times of up to $55\,\mathrm{Myr}$ for a GC distance of $150\,\mathrm{pc}$, but found that the spatial density of a young massive cluster drops quickly below the background density within a few Myr, rendering older clusters indetectable.

In spite of this difficulty, measuring the PDMF is essential to  deduce the IMF and to compare star formation in the GC with the outcome in less extreme environments. As the cluster centre is least affected by tidal stripping, as opposed to the outer cluster areas, the slope of the PDMF there can be compared to the PDMF slopes of spiral arm young massive clusters. In addition, tidally stripped stars might still be located close to the cluster at these young ages. If these stars can be identified as former cluster members, e.g. as they are co-moving with the cluster at comparable velocities, the common PDMF of cluster members and former members might still be a reasonable representation of the IMF. 
Most notably the measured PDMF of a GC cluster is an indispensable ingredient to constrain dynamical simulations set out to reconstruct its dynamical history, its IMF and to recover possible IMF variations. 

The name of the Quintuplet cluster arises from a "quintuplet" of bright near-infrared (NIR) sources (Q1-Q4,Q9), which was first observed by \citet{Nagata1990} and \citet{Okuda1990} by resolving one or two merged sources present in earlier surveys \citep[e.g.][]{Kobayashi1983}. Shortly afterwards, the number of resolved cluster stars was increased to 15 stars (Q1-Q15, \citet{Glass1990}, the "Q"-label was introduced in \citet{Figer1999}).
Until today,  21 Wolf-Rayet (WR) stars (14~WC, 7~WN stars, see \citealp[][and references therein]{Hucht2006}; \citealp{Mauerhan2010a,Liermann2009,Liermann2010}), two luminous blue variables (LBVs) and 93  OB stars were spectroscopically identified \citep{Figer1999,Liermann2009}. Hence, the Quintuplet hosts almost a quarter of the 92 WR stars in the Galactic centre region \citep{Mauerhan2010}.
The age of the cluster was derived based on spectral types of likely cluster stars to be $4 \pm 1\,\mathrm{Myr}$ \citep{Figer1999}. 
The Quintuplet cluster is therefore slightly older than the Arches cluster (age: $2-2.5\,\mathrm{Myr}$, \citealt{Najarro2004}), with whom it shares a similar projected GC distance ($30\,\mathrm{pc}$ and $26\,\mathrm{pc}$), a similar location in the sky and comparable masses on the order of $10^4\,\mathrm{M_{\sun}}$. As the density of the Quintuplet cluster is more than two orders of magnitude smaller than the density of the Arches cluster \citep{Figer2008}, and it exposes a much more dispersed appearance, it is suggestive to regard these two clusters as snapshots at different timesteps of massive clusters rapidly dissolving in the GC tidal field.

Due to the strong interstellar extinction for lines of sight towards the GC and the large distance to these clusters, NIR observations with high spatial resolution provided either by ground based telescopes with AO correction or space telescopes are required to study their stellar populations. As the Quintuplet cluster is far less compact than the Arches cluster, it is mandatory even in the cluster core to apply an effective method to discern the cluster stars from field stars in order to determine cluster properties in an unbiased way. 
In this paper, we analyse adaptive optics (AO) observations of the Quintuplet cluster obtained with the NAOS-CONICA (NACO) instrument on the ESO Very Large Telescope (VLT). A clean sample of cluster stars was established based on the measurement of individual proper motions and a subsequent colour-cut. The PDMF of the  central part of the Quintuplet cluster is derived from this cluster sample for the first time.

In Sect.~\ref{s_observations_and_data_reduction} we present the data and the data reduction. The photometric calibration and the determination of the astrometric and photometric uncertainties are detailed in Sect.~\ref{s_photometric_calibration_and_error_estimation}. The completeness of detected stars is derived from artificial star experiments in Sect.~\ref{s_completeness}. The proper motion of individual stars is determined in Sect.~\ref{s_proper_motion_membership}, and the proper motion membership sample is established. The colour-magnitude diagrams (CMDs) of proper motion members and non-members are discussed in Sect.~\ref{s_colour_magnitude_diagrams}, and the final cluster sample is derived applying a colour-cut and using spectral identifications from the spectral catalogue by \citet{Liermann2009}. Initial stellar masses are determined  in Sect.~\ref{s_mass_derivation} from four isochrones of different ages and different stellar models. The arising PDMFs of the four isochrones are discussed in Sect.~\ref{s_mass_functions}. In Sect.~\ref{s_discussion}, we compare our results to reported mass function slopes of other  young massive clusters in the Milky Way. This paper concludes with a short summary in Sect.~\ref{s_summary_outlook}.

\section{Observations and data reduction}
\label{s_observations_and_data_reduction}

\begin{figure}
  \resizebox{\hsize}{!}{\includegraphics{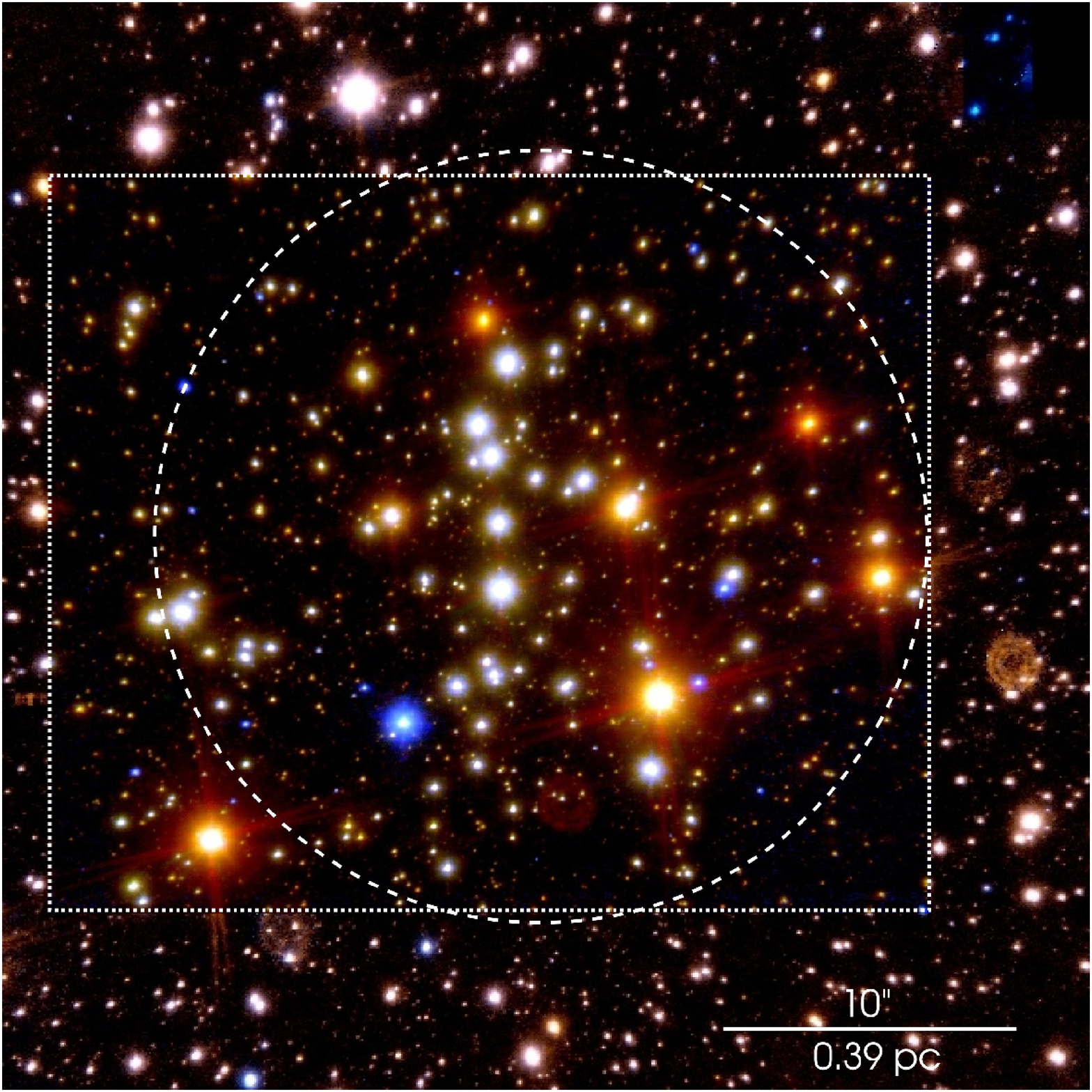}}
  \caption{VLT NACO $JHK_s$ composite image of the Quintuplet cluster. Outside the dotted rectangle only $H$- and $K_s$-band data are available. The dashed circle with a radius of $500\,\mathrm{pixel}$ or $0.5\,\mathrm{pc}$ indicates the region used for the derivation of the mass function (see Sect.~\ref{ss_data_selection_and_combination}). Due to bad AO correction, the $J$-band dataset was unsuitable to perform photometry and astrometry and was used only for this composite image.}
  \label{f_JHK_cluster}
\end{figure}

Two epochs of high-resolution observations of the Quintuplet Cluster ($\alpha = 17^\mathrm{h}46^\mathrm{m}15^\mathrm{s}$, $\delta=-28\degr49\arcmin41\arcsec$, J2000) were obtained in service mode on July 22-23th 2003 and July 24th 2008 with the Very Large Telescope (VLT), yielding a time baseline of $5.0\,\mathrm{yr}$. In order to achieve the high astrometric accuracy crucial for this study, the near-infrared imaging camera CONICA was used in combination with the NAOS instrument providing adaptive optics (AO) correction \citep{Lenzen2003,Rousset2003}. 
The first epoch was observed with a Santa Barbara Research Center (SBRC) InSB Aladdin2 array as detector, which was replaced by an InSB Aladdin3 array in May 2004 \citep{Nacoman}. All data were taken with the medium-resolution camera (S27) with a pixel scale of $0.0271\mathrm{\arcsec pixel^{-1}}$ and a field of view of $27.8\mathrm{\arcsec}$. The Quintuplet star Q2 \citep{Glass1990} with $K_s\sim6.6\,\mathrm{mag}$ served as the natural guide star for the infrared wavefront sensor. The data of the first epoch was retrieved from the ESO archive (PI: F.~Eisenhauer) and consisted of three datasets: two datasets in the $H$- and $K_s$-band  with a short detector integration time (DIT) of $2.0\,\mathrm{s}$ and NDIT (number of integrated DITs) of 30 and a further $K_s$-band dataset with a longer DIT of $20.0\,\mathrm{s}$ and NDIT of 2 to expand the range of observed magnitudes towards fainter stars. Each of these three datasets consists of 16 dithered frames resulting in an effective field of view of about $40\mathrm{\arcsec}\times40\mathrm{\arcsec}$. The total integration time at the central part of the observed field with maximum overlap was $16\,\mathrm{min}$ for the two datasets with the shorter DIT and $11\,\mathrm{min}\,40\,\mathrm{s}$ for the dataset with the longer DIT. The $K_s$-band observations of the second epoch in 2008 cover the same field of view as the first epoch, but were carefully designed to provide high astrometric accuracy. The DIT of this dataset was $2.0\,\mathrm{s}$ with NDIT of 15, resulting in a total integration time of $16\,\mathrm{min}\,30\,\mathrm{s}$ for the 33 frames used for the combined image out of a total of 44 frames. The properties of all four datasets are summarized in Table~\ref{t_observations} and a composite colour image is shown in Fig.~\ref{f_JHK_cluster}.

Standard data reduction was performed using a custom made reduction pipeline written in Python, calling a series of self-written IDL routines as well as PyRAF tasks. Both sky and science frames were reduced by subtracting the appropriate darks and dividing by the twilight flat fields.  One master sky was created per dataset with the PyRAF/IRAF task \emph{imcombine} by determining the average of the 2nd to 5th faintest pixel. All brighter pixels were rejected to avoid contamination with stellar light in the master sky image. This is particularly crucial along the line of sight towards the GC, where the high stellar density prohibits the choice of star-free sky fields.
For the 2008 dataset, the master sky was derived from both the sky and science frames as otherwise strong stellar residua remained resulting from too small dithers of the sky frames. Before the master sky image was subtracted from a science frame it was scaled to the background level of this frame. 
Hot and dead pixels were detected from outliers during the combination of the dark images and twilight flats and written to the master bad pixel mask. For each science frame, pixels affected by cosmic ray hits were determined by the IRAF task \emph{cosmicrays} and added to the master bad pixel mask to create an individual mask for each frame.

All data from 2008 except for the twilight flat fields were affected by the sporadic so-called $50\,\mathrm{Hz}$ noise from the Aladdin3 detector causing horizontal stripes, the intensity and position of which vary with time \citep[chap.~5.1]{Nacopipeman}. To correct for this noise, an appropriate routine from the ESO ECLIPSE pipeline \citep{Devillard2001} for the ISAAC instrument was rewritten in IDL and adapted to the NACO data.
It encompasses the following steps: 1.)~Determine the median of each row after rejecting the 40 darkest and the 420 brightest pixels in order to exclude bad pixels and flux from stars, and store the median values in an array. 2.)~Smooth the array of median values with a median filter with a  half-width of 40 pixel. 3.)~Subtract the difference between the original and the smoothed array of median flux values from each column of the image. This correction routine was applied to each reduced sky frame before they were combined to the master sky, and to each science frame after the subtraction of the corrected sky. The application of this routine after the basic reduction ensures that features due to dark current, detector bias and response are not altered by the $50\,\mathrm{Hz}$ noise correction, and are correctly removed using the dark and the twilight flat.  

Very bright sources cause a regular pattern of electronic ghosts, the sizes and shapes of which depend on the detector array and the brightness of the inducing object. The minimum stellar peak flux in the raw images sufficient to generate a visible electronic ghost was about $2700\,\mathrm{counts}$ for all three $2.0\,\mathrm{s}$ DIT datasets. For the dataset with the longer DIT of $20.0\,\mathrm{s}$ the minimum flux was about $5500\,\mathrm{counts}$, as ghosts of fainter stars are probably hidden within the brighter background. The location of the object on the detector array determines the position of these ghosts. A bright star at the position (x,y) causes electronic ghosts at (x,1024-y), (1024-x, 1024-y) and (1024-x, y) \citep[chap. 4.7.1]{Nacoman}. For each science frame a mask covering the visible electronic ghosts was generated and added to the mask containing the bad pixels in the dark, the flat field and bad pixels arising from cosmic ray hits. A few optical ghosts, apparent as sets of concentric rings with a radius of about $40\,\mathrm{pixel}$ ($\hat = 1.1\mathrm{\arcsec}$), are caused by the bright stars as well \citep[Fig.~16]{Nacoman}. Unlike the electronic ghosts, the position of these ghosts seems to be fixed with respect to the observed star field, which prohibits a correction for these ghosts during the image combination, as they are placed on the same location in the combined image. As the brightness of stars, residing within an optical ghost, cannot be reliably determined these stars were later removed from all source catalogues. 

The science frames of a dataset were combined into one combined image with the IRAF task \emph{drizzle} \citep{Fruchter2002}. Each science frame was linearly weighted by the inverse of the full width at half maximum (FWHM) of an unsaturated, bright star present in all frames of this dataset in order to optimise the spatial resolution in the combined image. Pixels contained in the  bad pixel masks are rejected by \emph{drizzle} during the image combination. The large number of 44 exposures for the second epoch allowed for using only the 33 frames with a FWHM $ <  0.082\mathrm{\arcsec}$ in order to enhance the spatial resolution without losing stars at the faint end.

Stellar fluxes and positions were determined with the \emph{starfinder} algorithm \citep{Diolaiti2000}, which is designed for high precision astrometry and photometry on AO data of crowded fields. The point spread function (PSF) is derived empirically from the data by median superposition of selected stars after subtraction of the local background and normalization to unit flux. Using an empirical PSF is preferable for astrometric AO data, as the steep core and wide halo are not well reproduced by analytic functions. Stars whose peak fluxes exceed the linearity limit of the detector and are included in the list of stars for the PSF extraction are repaired by replacing the saturated core with a replica of the PSF, scaled to fit the non-saturated wings of the star\footnote{Stars, whose peak flux exceed the linearity limit of the detector are referred to as saturated for the remainder of this paper.}. 
Only if the saturated stars are repaired they are definitely detected and fitted by the algorithm, so that their contribution on the flux of neighbouring stars can be subtracted. This is of special importance for faint stars located within the halo of a saturated star in order to measure their fluxes precisely.
As the option for a spatial variation of the PSF due to uncorrected atmospheric modes for larger distances to the AO guide star is not yet implemented in the \emph{starfinder} code, a spatially constant PSF was used. In order to obtain a valid estimate for the PSF for the most parts of the image, preferably isolated, bright stars uniformly spread across the image were selected for the PSF extraction. All saturated stars were included in the list of PSF stars in order to be repaired. The total number of selected PSF stars and the number of saturated stars among them are listed in Table~\ref{t_PSF_extraction}. The comparably small number of saturated stars of the last dataset is due to the higher linearity limit of the Aladdin3 detector.

The simplification of a constant PSF across the whole image led to spatially varying PSF fitting residuals and in turn to small-scale zeropoint variations across the field. This is typical for AO data and is mostly a consequence of anisoplanatism at increasingly larger distances from the natural guide star. As the extracted PSF resembles an average of the different PSFs across the image, the variation of the residuals after PSF subtraction is not centred at the position of the guide star. It rather shows a radial trend also depending on the respective image segment.
In order to correct for these variations a spatially varying correction factor was determined from the flux ratio of the residual flux in the PSF subtracted image and the stellar flux within an aperture around the centroids of isolated stars. 
Only stars without neighbours brighter than 1\% of the star's flux were selected. The radius of the aperture was chosen to be either $0.08\mathrm{\arcsec}$, $0.14\mathrm{\arcsec}$ or $0.22\mathrm{\arcsec}$ in dependence of the stellar flux to make sure that on the one hand the bulk of the residual and stellar flux is included, while on the other hand the residual flux for the faint stars is not dominated by the background noise. 
The flux ratio was fitted in dependence of the distance to the image centre for segments of 45$\degr$ either by a constant offset or a small linear trend. 
The correction factor $f_{\mathrm{corr},i}(r)$ for the i$^{\mathrm{th}}$ segment, which is to be multiplied to the fluxes of all stars in this segment, follows from the fit of the flux ratio ${\cal R}_{\mathrm{fit},i}(r)$: 
\begin{equation}
\label{e_corr_factor}
  f_{\mathrm{corr},i}(r) = \frac{1}{1-{\cal R}_{\mathrm{fit},i}(r)}\,.
\end{equation}
The error of $f_{\mathrm{corr},i}(r)$ is derived from the fitting error of ${\cal R}_{\mathrm{fit},i}(r)$, which is $\Delta {\cal R}_{\mathrm{fit},i}(r) = \Delta c_i$ if the flux ratio was fitted by a constant offset $c_i$ and $\Delta {\cal R}_{\mathrm{fit},i}(r) = \sqrt{(r\Delta b_i)^2+(\Delta c_i)^2}$ if the flux ratio was fitted by a linear trend with ${\cal R}_{\mathrm{fit},i}(r) = b_i r+ c_i$:
\begin{equation}
\label{e_corr_factor_error}
\Delta f_{\mathrm{corr},i}(r) = \left(\frac{1}{1-{\cal R}_{\mathrm{fit},i}(r)}\right)^2 \Delta {\cal R}_{\mathrm{fit},i}(r) \,.
\end{equation}
This procedure resulted in the most consistent photometric calibration across the observed field.

\section{Photometric calibration and error estimation}
\label{s_photometric_calibration_and_error_estimation}

Reference sources for the photometric calibration were taken from the Galactic Plane Survey (GPS) \citep{Lucas2008}, a part of the UKIRT Infrared Deep Sky Survey (UKIDSS) \citep{Lawrence2007}. Magnitudes of stars within the  UKIDSS catalogue are determined from aperture photometry using an aperture radius of $1\mathrm{\arcsec}$  and are calibrated using the Two Micron All Sky Survey (2MASS) \citep{Skrutskie2006}. Data from the Sixth Data Release (DR6) for the Quintuplet cluster was retrieved from the UKIDSS archive \citep{Hambly2008}. For the NACO H-band data (2003) and the second epoch $K_s$-band dataset (2008), sources from the UKIDSS catalogue in the Quintuplet Cluster served as zeropoint reference.
For a set of calibration stars (29 in $H$-, 13 in $K_s$-band), which could unambiguously be assigned to calibrated sources in the UKIDSS catalogue, the individual zeropoints were  determined. Due to the high spatial resolution of the NACO data usually several fainter stars can be resolved within the UKIDSS $1\arcsec$ aperture around the calibration star. As these stars do contribute to the measured flux in the UKIDSS survey, the PSF-flux of all stars falling within a radius of $r = 1\mathrm{\arcsec}-0.5 \times \mathrm{FWHM_{PSF}}$, with $\mathrm{FWHM_{PSF}}$ being the FWHM of the extracted PSF of the respective NACO dataset, was added and compared to the magnitude of the respective star in the UKIDSS catalogue. The final zeropoint was then determined from the average of the  individual zeropoints of the calibration stars. The zeropoints of the two $K_s$-band datasets from the first epoch were determined subsequently using the calibrated second epoch data. 
No significant colour terms were found between the NACO $H$, $K_s$ and the UKIDSS $H$,$K$ filter systems. 

\begin{figure}
  \resizebox{\hsize}{!}{\includegraphics{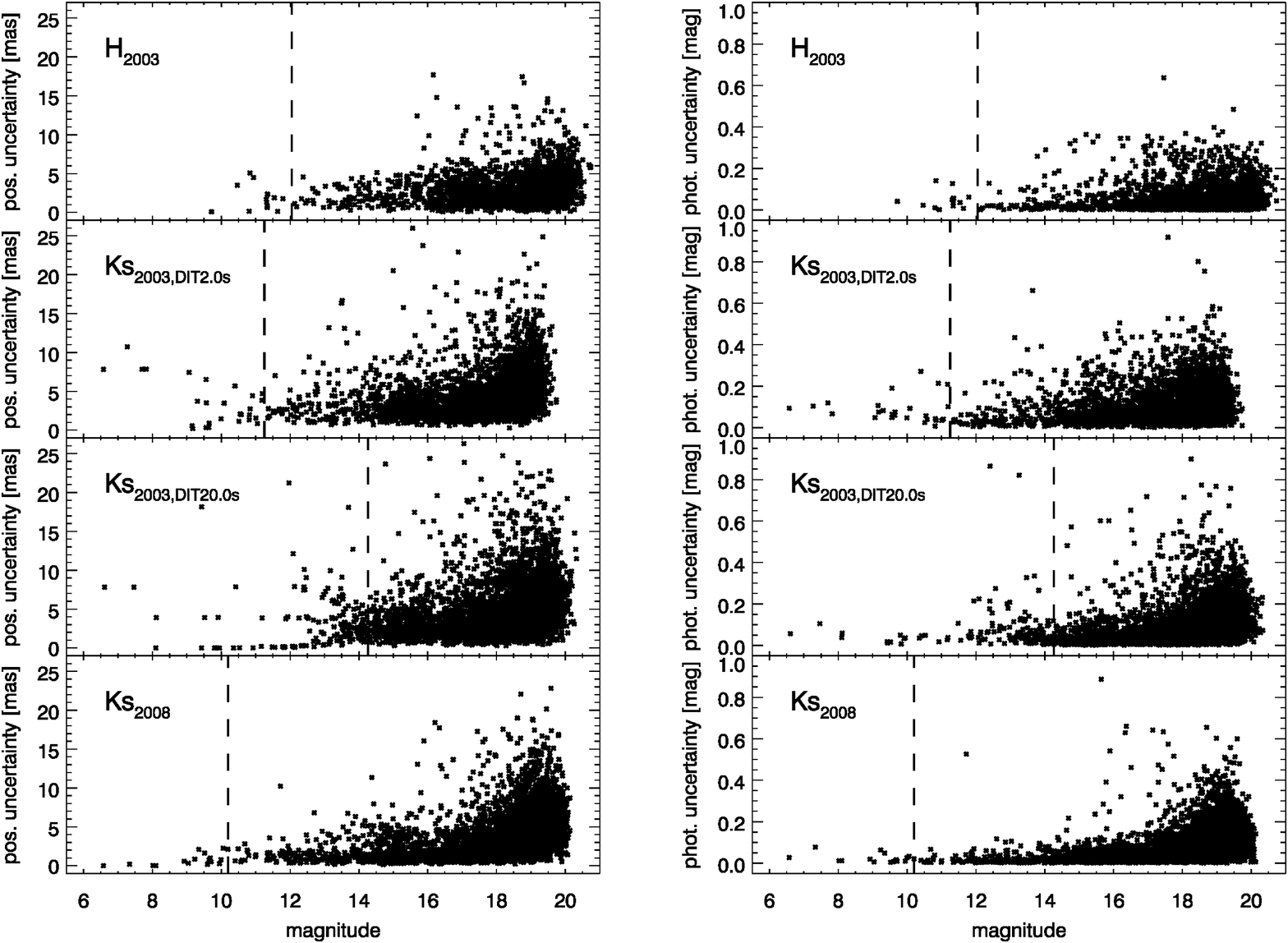}}
  \caption{Plot of the astrometric uncertainty (left panels) and the photometric uncertainty (right panels) vs. the magnitude for all four NACO datasets. The plotted photometric uncertainty does only include the PSF fitting uncertainty. The dashed lines mark the linearity limit of the respective dataset.}
  \label{f_POSERR_dataset_error_plots}
\end{figure}

The estimation of the photometric and astrometric uncertainties follows the approach described in \citet{Ghez2008} and  \citet{Stolte2008}. The reduced science frames for each dataset were divided into three subsets of comparable quality and coverage. Each subset of 5 (first epoch) or 11 frames (second epoch) was then combined with \emph{drizzle} and the photometry and astrometry of the resulting auxiliary image was derived with \emph{starfinder} in the same way  as for the deep images. Each auxiliary frame was calibrated with respect to the deep image. As we are only concerned about the three independent measurements no correction factor $f_{\mathrm{corr},i}(r)$ for the flux was applied, such that the photometric uncertainty derived from the auxiliary frames includes only the PSF fitting uncertainty.
The photometric and astrometric uncertainty was derived as the standard deviation of the three independent measurements for each star detected in all three auxiliary frames. As no preferential direction is expected for the positional uncertainty, the astrometric uncertainty of each star is computed as the mean of the positional uncertainty in the x- and y-direction. The astrometric and photometric uncertainties as derived from the auxiliary frames are shown in dependence of the magnitude in Fig.~\ref{f_POSERR_dataset_error_plots} for all datasets.

In order to remove false detections from the three $K_s$-band catalogues, only stars which were detected in all three auxiliary images of a dataset, and hence with measured astrometric and photometric uncertainties assigned to them, were kept in the respective source catalogue. 
For the $H$-band data this criterion was not applied. The $H$-band was matched in the further analysis (see Sect.~\ref{s_colour_magnitude_diagrams}) with a $K_s$-band catalogue containing only stars detected in both epochs. It is assumed that a star found in the $K_s$-band images of both epochs is a real source and if it is missing in one of the $H$-band auxiliary images this is a consequence of the substantially lower photometric depth of the auxiliary image.

The photometric errors as stated in the final source catalogue (Table~\ref{t_source_catalogue}) do include the respective zeropoint uncertainties, the photometric uncertainties due to the flux measurement from PSF fitting, and the error of the correction factors as given in equation~(\ref{e_corr_factor_error}). 

\begin{table*}
\caption{Overview of the VLT/NAOS-CONICA datasets.}
\label{t_observations}
\centering
\begin{tabular}{clccccccccc}
\hline\hline
Dataset No. & Date & Filter & No. of Frames & DIT & NDIT & $\mathrm{t_{int}}$\tablefootmark{a} & Airmass & Seeing& FWHM of PSF & Strehl ratio\\
&     &	      &		      & $\mathrm{(s)}$ & & $\mathrm{(s)}$ & & ($\arcsec$) & ($\arcsec$) &  \\
1 & 2003-07-22	& $H$ & 16 	&	2.0	& 30	&	960 & 1.00--1.02 & 0.47--0.60 &	0.078 & 0.15\\
2 & 2003-07-22	& $K_s$ & 16 	&	20.0	& 2	&	640 & 1.03--1.06 & 0.36--0.49 &	0.080 & 0.22\\
3 & 2003-07-23	& $K_s$ & 16 	&	2.0	& 30	&	960 & 1.03--1.07 & 0.31--0.47 &	0.082 & 0.26\\
4 & 2008-07-24	& $K_s$ & 33 	&	2.0	& 15	&	990 & 1.00--1.01 & 0.49--0.60 &	0.080 & 0.26\\
\end{tabular}
\tablefoot{
  \tablefoottext{a}{Total integration time of the central part of the image with maximum overlap.}
}
\end{table*}

\begin{table}
\caption{Number of stars for PSF extraction.}
\label{t_PSF_extraction}
\begin{tabular}{clcc}
\hline\hline
Dataset No. & No. of PSF stars & No. of saturated PSF stars\\
1  & 37 & 17 \\
2  & 239 & 136 \\
3  & 48 & 29 \\
4  & 69 & 15 \\
\end{tabular}

\end{table}

\section{Completeness}
\label{s_completeness}

In order to quantify the detection losses due to crowding effects, the local completeness for each dataset was determined from the recovery fraction of artificial stars inserted into each combined image.
The artificial star experiment for the $H$-band data covers a magnitude range from $9.5$ to $21.5\,\mathrm{mag}$. For each magnitude bin with a width  of $0.5\,\mathrm{mag}$, 42~artificial star fields were generated. Each artificial star field was created by adding 100~artificial stars, which are scaled replica of the empirical PSF, inserted at random positions and with random fluxes within the respective flux interval, into the combined image.

\begin{figure}
 \resizebox{\hsize}{!}{\includegraphics[angle = 90]{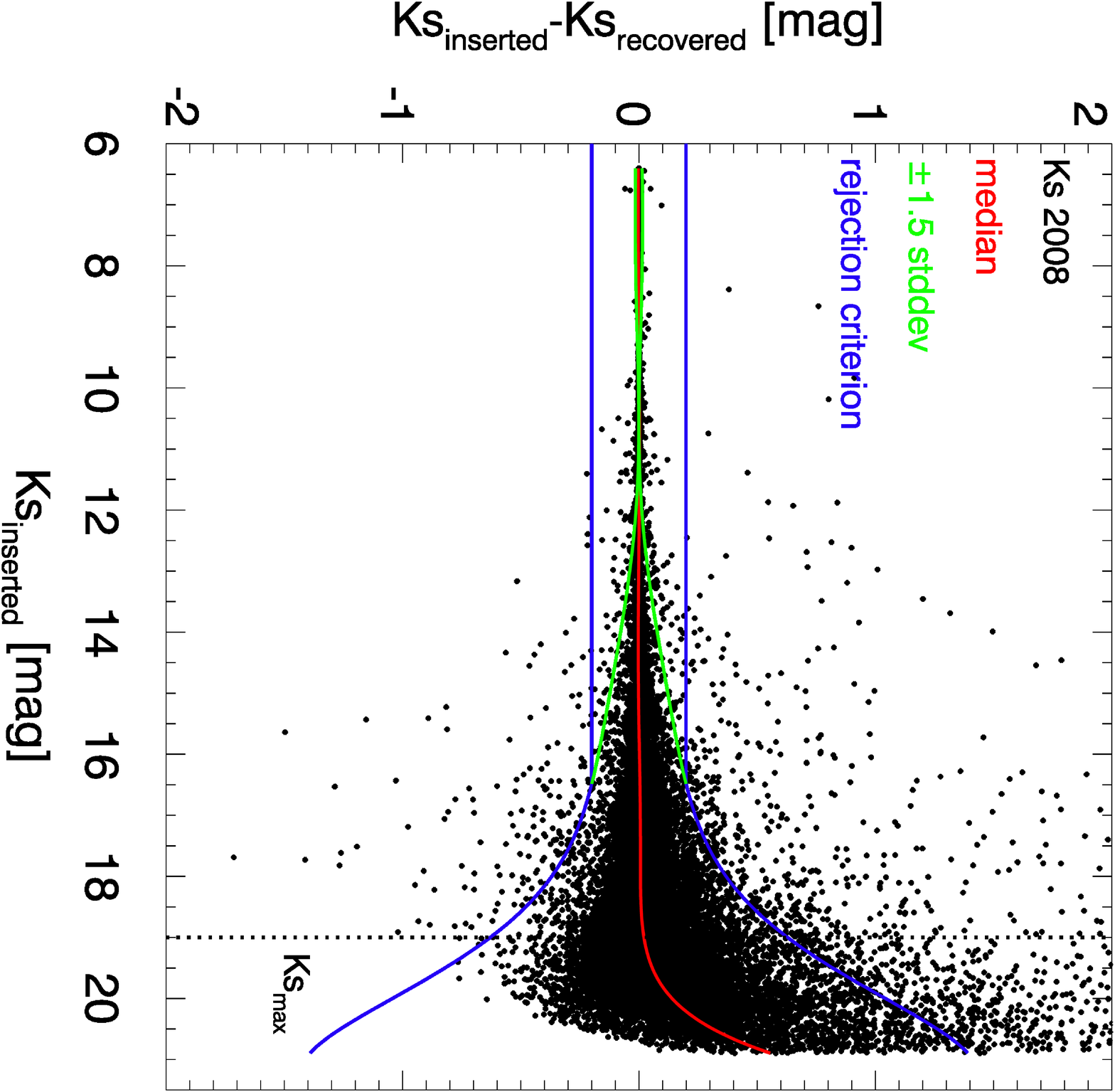}}
 \caption{Difference of the inserted and recovered magnitudes of artificial stars inserted into the combined image of the $K_s$-band data in 2008 plotted in dependence of the magnitude. A high-order polynomial fit to the median and the standard deviation (multiplied by a factor of 1.5) of the magnitude difference within magnitude bins of $1\,\mathrm{mag}$ are shown as well. If the absolute magnitude difference exceeds $0.20\,\mathrm{mag}$ and is larger than $1.5$ times the fit to the standard deviation, a recovered artificial star is rejected and treated as non-recovered.
 The vertical dotted line indicates the maximum $K_s$-band magnitude at $K_s = 19\,\mathrm{mag}$ of stars to be used for the proper motion analysis.}
 \label{f_incompleteness_magdiff}
\end{figure}

For the three $K_s$-band datasets, the artificial stars were inserted at the same physical positions as in $H$-band and with a magnitude in $K_s$ yielding a colour for the respective artificial star of $H-K_s = 1.6\,\mathrm{mag}$, which resembles the colour of main sequence (MS) stars in the Quintuplet cluster (see Sect.~\ref{s_colour_magnitude_diagrams}). 
The photometry on the images with added artificial stars was performed in the same way and using the same PSF as for the original images. In addition to artificial stars which were not re-detected by \emph{starfinder}, also stars where the recovered magnitudes deviated strongly from the inserted magnitudes, were considered as not recovered. The criterion to reject recovered stars due to their magnitude difference between input and output magnitude was derived from polynomial fits to the median and the standard deviation of the magnitude difference within magnitude bins of $1\,\mathrm{mag}$ (Fig.~\ref{f_incompleteness_magdiff}). Extreme outliers with absolute magnitude differences larger than 20 times the standard deviation were excluded from the determination of the median and standard deviation used for the fits. Stars with absolute magnitude differences exceeding $0.20\,\mathrm{mag}$ and being larger than $1.5$ times the fit to the standard deviation are treated as not recovered. 
The median of the magnitude difference exposes a systematic increase towards the faint end, exceeding $0.05\,\mathrm{mag}$ for  $K_s > 19.4\,\mathrm{mag}$ or $H > 20.25\,\mathrm{mag}$. This trend indicates that for the faintest stars the measured fluxes are not reliable anymore. As we restrict the analysis to stars brighter than $K_s < 19\,\mathrm{mag}$, sources at these faint magnitudes are excluded from the proper motion and mass function derivation.

\begin{figure*}
 \resizebox{\hsize}{!}{\includegraphics{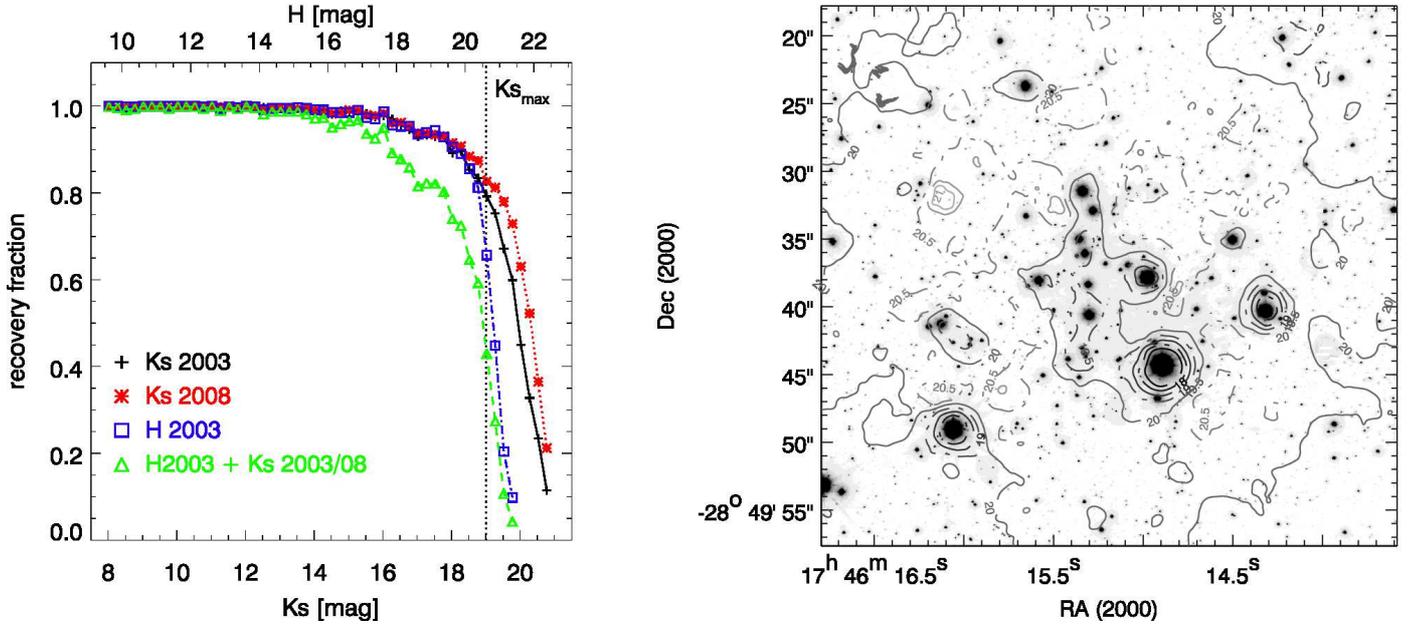}}
 \caption{Left panel: Recovery fractions of artificial stars inserted within the inner $500\,\mathrm{pixel}$  from the centre of the observed field plotted vs. the respective magnitude in the $K_s$- (lower abscissa) or $H$-band (upper abscissa). The position of the artificial stars were identical in the $K_s$- and $H$-band images and a formal colour of $H-K_s = 1.6\,\mathrm{mag}$ was assigned to each star. The full and dotted lines correspond to the recovery fractions for the $K_s$-band data in 2003 and 2008, respectively and the dash-dotted line shows the completeness in the $H$-band. The dashed line shows the total completeness for the stars after matching the two $K_s$-band and the $H$-band datasets. Only stars with $ K_s < 19\,\mathrm{mag}$ are used for the proper motion analysis, as indicated by the vertical dotted line. Right panel: $K_s$-band image from the second epoch with the overplotted contours representing  a completeness level of 50\% for the labelled magnitudes.}
 \label{f_incompleteness_incompleteness}
\end{figure*}

The left panel in Fig.~\ref{f_incompleteness_incompleteness} shows the overall recovery fraction in the $K_s$-band for the two epochs and the $H$-band within a radius of $500\,\mathrm{pixels}$ ($\hat = 13.6\mathrm{\arcsec}$) from the image centre, the part of the image actually used for the determination of the present-day mass function (see Sect.~\ref{s_mass_functions}). The shown recovery fraction for the $K_s$-band  data from 2003 is a combination of the recovery fractions for the two $K_s$-band datasets of that epoch. The dataset with the longer DIT of $20.0\,\mathrm{s}$ is used only for magnitudes fainter than the linearity limit of this dataset at $14.3\,\mathrm{mag}$. For brighter magnitudes, the recovery fraction of the 2003 $K_s$-band data with the short DIT of $2.0\,\mathrm{s}$ is drawn. The star catalogues of these two datasets are combined accordingly in the subsequent analysis (see Sect.~\ref{ss_data_selection_and_combination}).
The total recovery fraction also shown in the figure is the product of all three recovery fractions and is most relevant for the completeness correction of the mass function, as i) only stars which are detected in both epochs can be proper motion members and ii) only for stars with measured $H$-band magnitudes can masses be derived. 

Completeness varies as a function of position due to the non-uniform distribution of brighter stars and hence is a function of the stellar density and magnitude contrast between neighbours \citep[see e.g.,][]{Eisenhauer1998,Gennaro2011}. A spatially-dependent approach to determine the local completeness value becomes especially important if the cluster exhibits a non-symmetric geometry or in the presence of sparse very bright objects, which heavily affect the completeness values in their surrounding as in the case of the Quintuplet cluster. 
In order to assign a local completeness value to each detected star, the method described in Appendix~A of \citet{Gennaro2011} was applied to derive completeness maps for each combined image containing the recovery fraction for every pixel as a function of magnitude. The procedure encompasses threes steps performed for each magnitude bin (for a detailed description the reader is referred to \citealt{Gennaro2011}): 1.)~Derive the local, averaged completeness value at the position of each artificial star from the number of recovered nearest neighbours including the star itself. 2.)~Interpolate these local completeness values into the regular grid of image pixels. 3.)~Smooth the obtained map with a boxcar kernel with a width of the sampling size in order to remove potential artificial features introduced by the previous step. 
For the last step the completeness maps of all magnitude bins are used. To ensure that the completeness decreases monotonically with increasing magnitude, a Fermi-like function is fitted to the completeness values at every pixel in the image as a function of magnitude. The completeness (or recovery fraction) for every real star in the respective band and dataset can then be computed from the fit parameters at the position of the star.
The right panel in Fig.~\ref{f_incompleteness_incompleteness} shows the combined $K_s$-band image for the second epoch with superimposed $50\%$-completeness contours stating the limiting magnitudes. The very bright stars with their extended halos hamper the detection of nearby faint stars causing the recovery fraction to be non-uniform across the field, as expected.
The completeness of a star entering the mass function is the product of its completeness in the $H$-band, the second epoch $K_s$-band data and either in the $2.0\,\mathrm{s}$ DIT ($K_{s,2003} < 14.3\,\mathrm{mag}$) or the $20.0\,\mathrm{s}$ DIT ($K_{s,2003} > 14.3\,\mathrm{mag}$) first epoch $K_s$-band dataset as determined from the respective completeness maps: 
\begin{equation}
f_{\mathrm{comp}} = f_{\mathrm{comp,Ks2008}} \times f_{\mathrm{comp,Ks2003}} \times f_{\mathrm{comp,H2003}}\,.
\end{equation}
For stars brighter than $H = 13.5\,\mathrm{mag}$ or $K_s = 10.4\,\mathrm{mag}$ the completeness was assumed to be $100\%$.

\section{Proper motion membership}
\label{s_proper_motion_membership}

Due to the high field star density for lines of sight towards the Galactic centre the distinction between cluster and field stars becomes particularly important. As most of the field stars are located within the Galactic bulge they have similar extinction values as the cluster and cannot be distinguished from cluster members on the basis of their colours alone. 
The high astrometric accuracy of the AO assisted VLT observations in combination with the time baseline of $5.0\,\mathrm{yr}$  allows  for the measurement of the individual proper motions of stars at the distance of the Quintuplet cluster. The primary applied method to discern the cluster members from the field stars is based on the measured proper motions.

\subsection{Geometric transformation}
\label{ss_geometric_transformation}

In order to determine the spatial displacements, two geometric transformations were derived to map each position in the two first epoch $K_s$ images (2003) with short ($2.0\,\mathrm{s}$) and long ($20.0\,\mathrm{s}$) DIT onto the correct position in the second epoch $K_s$ image (2008). 
The second epoch is used as reference epoch because of the higher astrometric accuracy, deeper photometry and brighter linearity limit of this dataset. 
Only the $K_s$-band datasets were used to determine the spatial displacements, as due to their higher Strehl ratios the stellar cores are better resolved  than in $H$-band, providing the better centroiding accuracy and hence most accurate astrometry.  

Under the assumption that internal motions are not resolved so far, the cluster itself served as the reference frame. 
The geometric transformation was derived in an iterative process. First, a rough transformation was determined with the IRAF task \emph{geomap} using the positions of manually selected bright, non-saturated stars uniformly distributed across the images of both epochs. The respective catalogue of the first epoch dataset (2003) was then mapped onto the catalogue of the second epoch (2008) to get a mutual assignment of stars found in both catalogues.
From these stars the most likely cluster candidates were selected to provide the reference positions for the refined, final geometric transformation. As the bright stars used for the preliminary transformation are likely cluster members the distribution of spatial displacements in the x-,y-direction of cluster star candidates are expected to scatter around the origin. Therefore only stars with spatial displacements within a radius of $0.8\,\mathrm{pixel} \hat= 4.3\,\mathrm{mas/yr}$ from the origin were selected for the derivation of the final transformation, which excludes most of the presumed field stars. Further, as the bulk of cluster stars are probably brighter than most stars in the field, only non-saturated bright and intermediate bright stars ($11.5<K_s<15.5\,\mathrm{mag}$ for a DIT of $2.0\,\mathrm{s}$ and $14.0<K_s<17.0\,\mathrm{mag}$ for a DIT of $20.0\,\mathrm{s}$) provide the reference positions.
The final geometric transformations were derived with \emph{geomap} in an interactive way. The residual displacements in the x-, y-directions between the transformed first epoch and the second epoch coordinates were minimized by iteratively removing outliers and carefully adapting the order of the polynomial fit ($=3$ for the final transformations). 
The final rms deviation of the geometric transformation was $0.2\,\mathrm{mas/yr}$ in the x- and $0.3\,\mathrm{mas/yr}$ in the y-direction for the dataset with a DIT of $2.0\,\mathrm{s}$, and $0.3\,\mathrm{mas/yr}$ in the x- and y-direction for the dataset with a DIT of $20.0\,\mathrm{s}$.
For a total cluster mass of $M_{cl} \approx 5900\,\mathrm{M_{\sun}}$ within a radius of $r\leq0.5\,\mathrm{pc}$ (see Sect.~\ref{s_mass_functions}) the internal velocity dispersion is expected to be on the order of $0.15-0.2\,\mathrm{mas/yr}$ or $6-8\,\mathrm{km/s}$. As this is smaller than the uncertainty of the geometric transformation alone, intrinsic motions are not resolved. Therefore the above assumption is justified.

\subsection{Data selection and combination}
\label{ss_data_selection_and_combination}

\begin{figure*}
  \resizebox{\hsize}{!}{\includegraphics{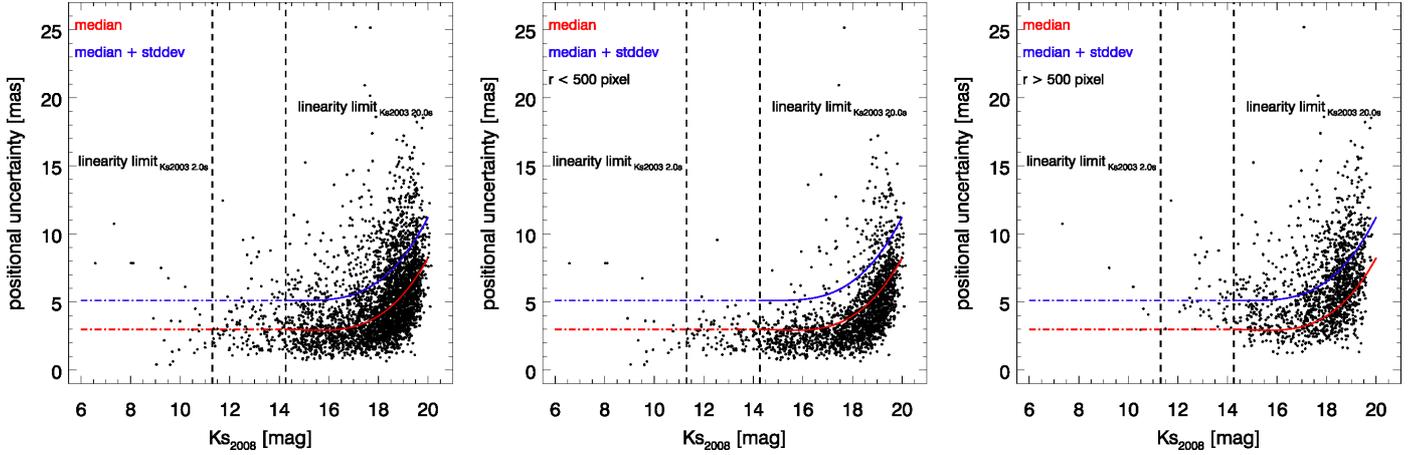}}
  \caption{Left panel: Plot of the combined astrometric uncertainty from the two epochs of $K_s$-band data plotted vs. the magnitude of the second epoch (for details of the error estimation see Sect.~\ref{s_photometric_calibration_and_error_estimation}). 
  The median and standard deviation of the astrometric uncertainty above the linearity limit (at $14.3\,\mathrm{mag}$) of the dataset in 2003 with $20.0\,\mathrm{s}$ DIT were fitted by polynomials. This fit of the median (lower line) and the sum the median and standard deviation (upper line) are drawn in all three plots. 
  Middle panel: Astrometric uncertainty of stars residing within a circle with $r < 500\,\mathrm{pixel}\,\hat =\, 13.6\arcsec$ around the centre of the observed field.
  Right panel: Astrometric uncertainty of stars residing outside this radius, which are excluded from further analysis.
  }
  \label{f_POSERR_match_ks_2008_2003_poserr_flag}
\end{figure*}

Each of the two transformed star catalogues of the $K_s$-band data from the first epoch was matched with the star catalogue of the second epoch using a matching radius of $4\,\mathrm{pixels}$ ($\hat=108\,\mathrm{mas} = 1.4 \times \mathrm{FWHM_{PSF}}$). The matching radius was chosen small enough to avoid mismatches between close neighbouring stars, but large enough to include all moving sources at GC distances below the escape velocity of the GC. A displacement of $108\,\mathrm{mas}$ within the time baseline of $5.0\,{\mathrm{yr}}$ for a distance of $8.0\,\mathrm{kpc}$ to the GC \citep{Ghez2008} corresponds to a proper motion of $820\,\mathrm{km/s}$.
The combined astrometric uncertainty $\sigma_{\mathrm{pos}} = \sqrt{\sigma_{\mathrm{pos,Ks2003}}^2+\sigma_{\mathrm{pos,Ks2008}}^2}$ was derived for both of these catalogues. In the left panel of Fig.~\ref{f_POSERR_match_ks_2008_2003_poserr_flag}, the combined astrometric uncertainty is plotted against the magnitude. The datapoints below the linearity limit of the long exposure in 2003 at $K_s = 14.3\,\mathrm{mag}$ originate from the  match of the second epoch with the data with a DIT of $20.0\,\mathrm{s}$, the datapoints at brighter $K_s$ magnitudes are from the match with the data obtained with a shorter DIT of $2.0\,\mathrm{s}$. 
For the matched catalogue using the first epoch dataset with a DIT of $20.0\,\mathrm{s}$, the median and the standard deviation of the astrometric uncertainties within bins of $0.5\,\mathrm{mag}$ width were fitted by a third and second order polynomial, respectively. The fit to the median and the sum of both fits are shown in all three panels of Fig.~\ref{f_POSERR_match_ks_2008_2003_poserr_flag} for comparison.
The usage of one averaged PSF for the whole image results in the observed radial increase of the PSF fitting residuals (see Sect.~\ref{s_observations_and_data_reduction}). The centroiding accuracy is therefore expected to decrease towards larger radii resulting in a larger astrometric uncertainty. 
The centre and right panel of Fig.~\ref{f_POSERR_match_ks_2008_2003_poserr_flag} exemplify  this behaviour by using only stars with a distance of less than or greater than $500\,\mathrm{pixel}$ $\hat=\, 13.6\arcsec$ from the centre of the combined images in both epochs, respectively. The decrease in the scatter and magnitude of the astrometric uncertainties is striking. The median of the astrometric uncertainty for $12 < K_s < 18 \,\mathrm{mag}$ is $2.52\,\mathrm{mas}$ for stars within a radius of $500\,\mathrm{pixel}$, but $4.58\,\mathrm{mas}$ for stars outside that radius.
Therefore, the further analysis is restricted to stars within a radius of $13.6\arcsec$ from the centre of the observed field of view for the remainder of this paper. The astrometric uncertainties rise steeply near the detection limit at about $20\,\mathrm{mag}$ (see centre panel in Fig.~\ref{f_POSERR_match_ks_2008_2003_poserr_flag}). For stars fainter than $19\,\mathrm{mag}$,  almost no stars exhibit an uncertainty below the median value of stars with intermediate brightness ($14<K_s<17\,\mathrm{mag}$). Stars with a $K_s$-band magnitude fainter than $19\,\mathrm{mag}$ are therefore excluded from the sample.
As last selection based on the combined uncertainty, stars fainter than $K_s = 14.3\,\mathrm{mag}$ are removed if their uncertainty is above the sum of the fits of the median and standard deviation derived from the combined uncertainty of all observed stars (see Fig.~\ref{f_POSERR_match_ks_2008_2003_poserr_flag}). The percentage of rejected stars varies between $0$ and $9.5\%$ for the affected magnitude bins and does not show a systematic trend with magnitude, therefore no systematic bias is introduced by this selection.   
After the above mentioned selections, the two matched catalogues were combined. Stars fainter than $K_s = 14.3\,\mathrm{mag}$ were taken from the match with the DIT $20.0\,\mathrm{s}$ first epoch data, brighter stars originate from the matched catalogue using the dataset from the first epoch with a DIT of $2.0\,\mathrm{s}$. 
The final $K_s$-band catalogue contains a total of 1297 stars. 

\subsection{The proper motion diagram}
\label{ss_the_proper_motion_diagram}

\begin{figure*}
  \resizebox{\hsize}{!}{\includegraphics{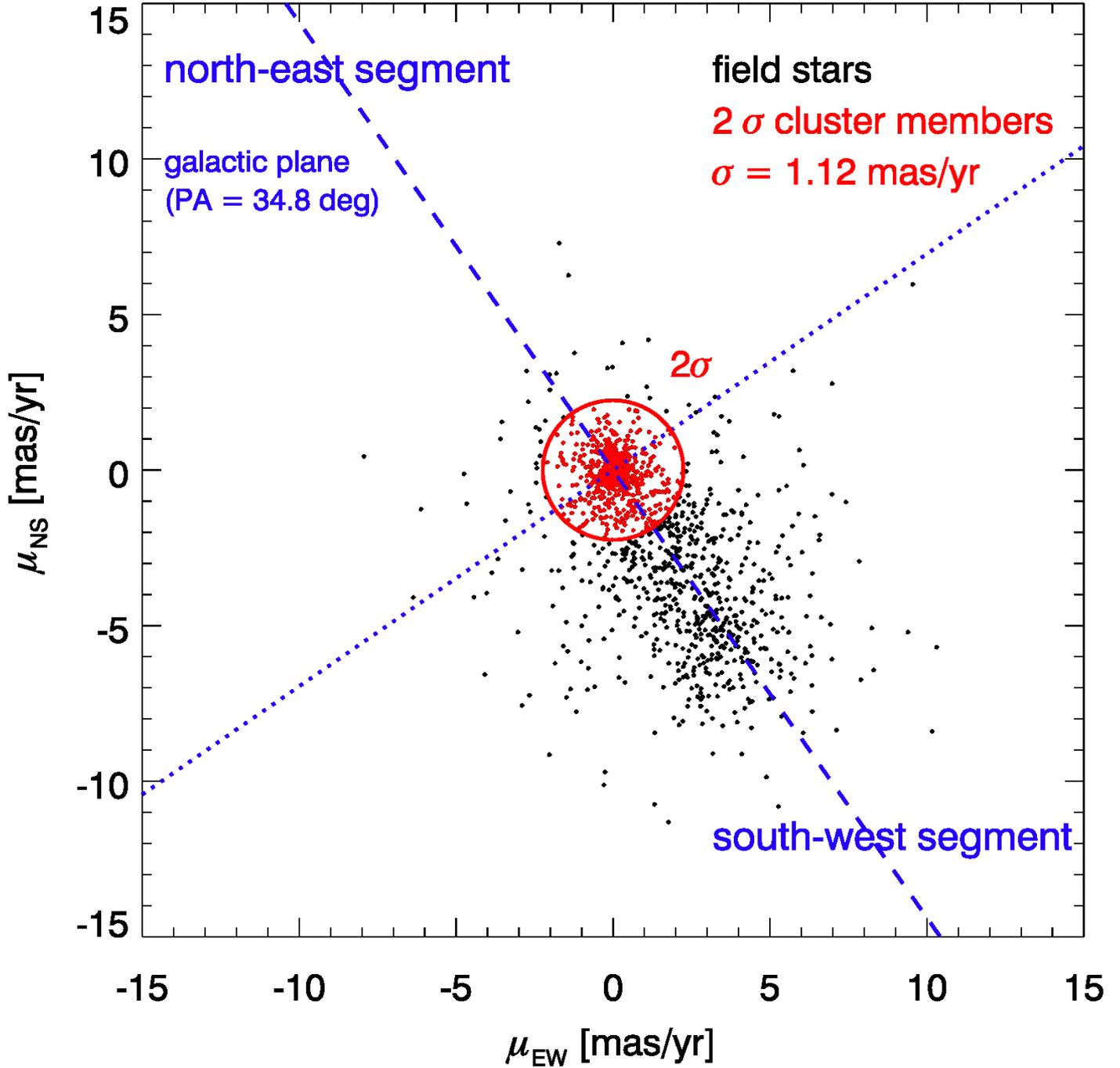}}
  \caption{Proper motion diagram of stars with $K_s \leq 19\,\mathrm{mag}$. The dashed line marks the direction parallel to the Galactic plane, the dotted line is oriented vertically with respect to the Galactic plane and splits the proper motion diagram into the North-East and the South-West segment. Stars within a radius of $2\,\sigma$ as derived from the Gaussian fit in Fig.~\ref{f_PMD_movehist} (right panel) around the origin are selected as cluster members.}  
  \label{f_PMD_member_selection}
\end{figure*}

Individual stars are plotted in the proper motion diagram (Fig.~\ref{f_PMD_member_selection}) with proper motions in the east-west-direction on the x- and proper motions in the north-south direction on the y-axis. As the cluster is used as the reference frame, the distribution of cluster members is centred around the origin and overlaps with the elongated distribution of the field stars. The orientation of the field stars is approximately parallel to the plane of the Galaxy (dashed line in Fig.~\ref{f_PMD_member_selection}). The dotted line running through the origin and vertically to the Galactic plane splits the proper motion diagram into two halfs being referred to as the North-East segment (upper half) and the South-West segment (lower half). 

\begin{figure*}
  \resizebox{\hsize}{!}{\includegraphics{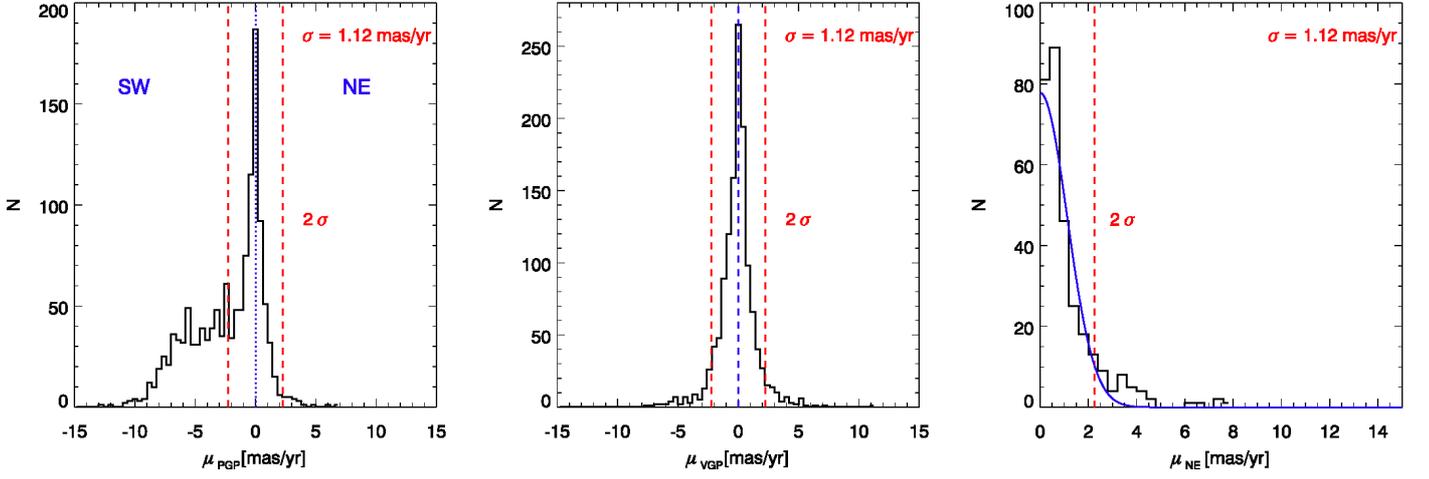}}
  \caption{Left panel: Histogram of proper motions parallel to the Galactic plane. The strongly peaked distribution of the proper motions for the cluster stars around the origin overlaps with the distribution of the fields stars located predominantly in the South-West-segment (see also Fig~\ref{f_PMD_member_selection}). Middle panel: Histogram of proper motions vertical to the Galactic plane. Right panel: Histogram of the 2-dimensional proper motions located in the North-East-segment of the proper motion diagram.  The histogram was fitted with a Gaussian function and a $2\,\sigma$ cut was used as the selection criterion for cluster membership (red dashed line in all three panels).}
  \label{f_PMD_movehist}
\end{figure*}

Figure~\ref{f_PMD_movehist} shows histogram plots of the distribution of the proper motions parallel (left panel) and vertical to the Galactic plane (centre panel). The distribution of proper motions in the direction parallel to the Galactic plane is strongly peaked at the origin, with a very steep decline in the North-West segment, and a slightly broadened decline and overlap with the broad field star distribution in the South-West segment, as expected from Fig.~\ref{f_PMD_member_selection}. The proper motions vertical to the Galactic plane are almost distributed symmetrically with respect to the Galactic plane (centre panel in Fig.~\ref{f_PMD_movehist}), confirming the assumed orientation of the field star distribution in the proper motion diagram. This and the exposed offset of the field star distribution in the proper motion diagram indicate a movement of the Quintuplet Cluster parallel to the Galactic plane towards  North-East with respect to the field as was found previously for the Arches Cluster \citep{Stolte2008}.
The sample of stars with proper motions in the North-East segment is least contaminated by field stars and was therefore used to derive the membership criterion. The distribution of proper motions in the North-East segment was fitted with a Gaussian function (right panel in Fig.~\ref{f_PMD_movehist}). Stars whose proper motions are within a circle of radius $2\,\sigma = 2.24\,\mathrm{mas/yr}$, where $\sigma$ is the width of the Gaussian fit, are selected as cluster members (see Fig.~\ref{f_PMD_member_selection}). Two of the initial five Quintuplet members (Q1, Q9) \citep{Nagata1990,Okuda1990} do not fall inside this circle. Their fluxes are exceeding the linearity limits by a factor of 8-30, such that their positions are not well determined. Note that this only affects the very brightest sources, for which spectroscopic member identification is available \citep{Figer1999,Liermann2009}. These two stars were added manually to the sample of proper motion members.

\section{Colour-magnitude diagrams}
\label{s_colour_magnitude_diagrams}

\begin{figure*}
  \resizebox{\hsize}{!}{\includegraphics{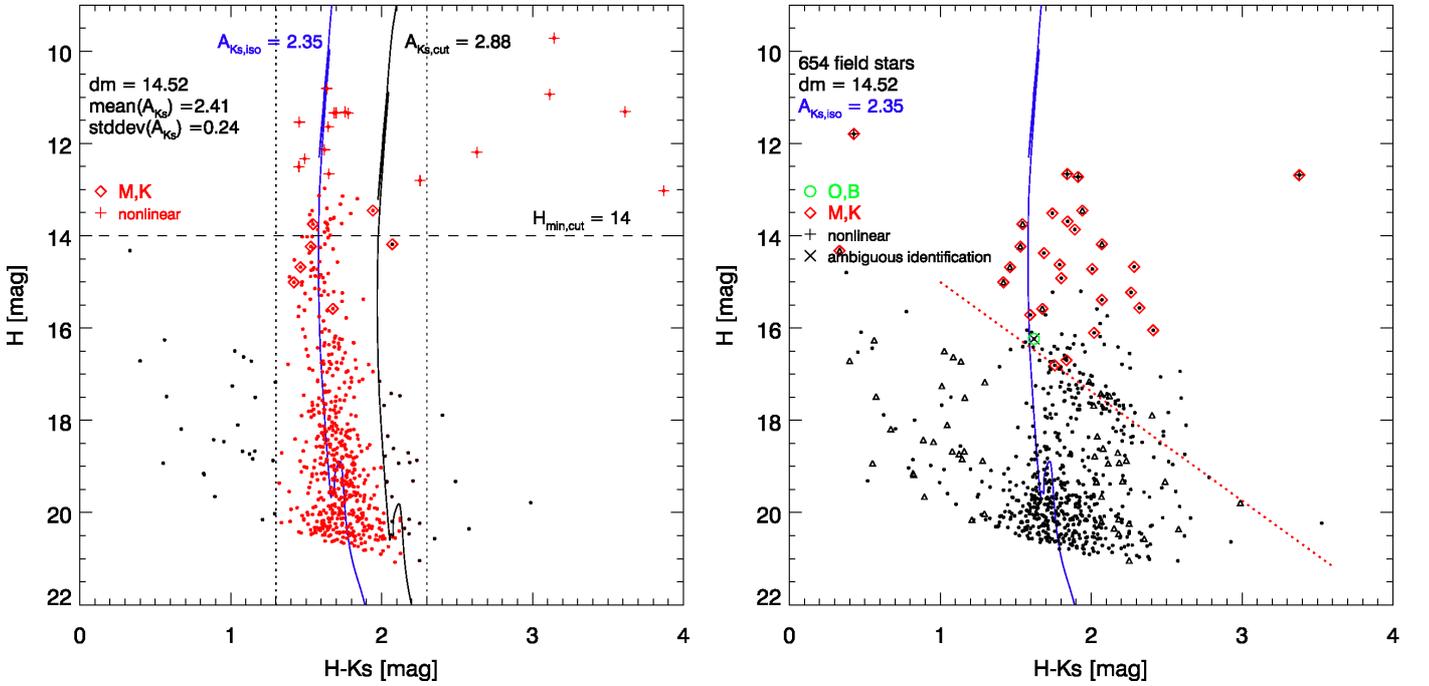}}
  \caption{Left panel: Colour-magnitude diagram of cluster member candidates on the basis of their proper motions. A $4\,\mathrm{Myr}$ isochrone  with solar metallicity, combined using a Padova MS-isochrone and a Pisa-FRANEC PMS-isochrone, shifted to a distance of $8\,\mathrm{kpc}$ and a foreground extinction of $A_{K_s} = 2.35\,\mathrm{mag}$, is shown for reference. Stars, with fluxes exceeding the linearity limit of the detector are drawn as crosses in all figures throughout this paper. 
  A small number of field stars is still present after the proper motion selection. These stars have similar proper motions as the cluster members, but most of them  can be distinguished on the basis of their colours. For this purpose, a two-step colour-cut was applied for stars with $H > 14\,\mathrm{mag}$ (see Sect.~\ref{s_colour_magnitude_diagrams} for details).
  The two short-dashed vertical lines mark the first colour-cut at $H-K_S = 1.3$ and $2.3$, rejecting blue fore- and red background stars. Highly reddened objects to the right of the second isochrone ($A_{K_s} = 2.88\,\mathrm{mag}$, black solid line) are removed as the second step of the colour-cut. The sample of cluster stars after the colour-cut is plotted in red, stars rejected based on their colour are drawn in black. 
  Spectroscopically identified field giants from the spectral catalogue of \protect{\citet{Liermann2009}} are marked with diamonds and are removed from the final cluster sample. 
  Right panel: Colour-magnitude diagram of stars classified as belonging to the field according to their proper motion (dots) and of stars removed from the member sample based on their colour or known spectral type (triangles). One star, classified as belonging to the field by its proper motion, has an O-star as (ambiguous) counterpart in the Liermann spectral catalogue (see Sect.~\ref{s_colour_magnitude_diagrams}) and is marked with a circle.
  The tilted dotted line is the line of reddening according to the extinction law by \protect{\citet{Nishiyama2009}} running through the population of red clump stars from the Galactic bulge.}
  \label{f_CMD_member_nonmember_cc}
\end{figure*}

The $K_s$ source catalogues of proper motion members and non-members were matched with the source catalogue of the first epoch H-band data.
All 1218 matched stars (member and non-members) are included in the final source catalogue (see Table~\ref{t_source_catalogue}).
The corresponding colour-magnitude diagrams (CMD) are shown in Fig.~\ref{f_CMD_member_nonmember_cc} and use only magnitudes in $H$ and $K_s$ from the first epoch to avoid additional scatter being introduced by variable stars. Stars whose fluxes exceed the respective linearity limit in either $H$ and/or $K_s$ at $H = 12.05\,\mathrm{mag}$ and $K_s = 11.25\,\mathrm{mag}$ are marked with crosses. 
All 1218 stars (member and non-members) with measured proper motion and $(H-K_s)$ colour are included in the final source catalogue (see Table~\ref{t_source_catalogue}).

Cluster and field stars separate well as can be seen by characteristic features of the field population (right panel), that are absent in the cluster selection (left panel). For example an elongated overdensity is observed, which  starts at about $H = 17\,\mathrm{mag}$ , $H - K_s = 1.8\,\mathrm{mag}$ and extends to redder colours along the reddening path adopting the extinction law by \cite{Nishiyama2009}. It is consistent with arising from red clump stars located in the Galactic bulge. Assuming the intrinsic $K$-band magnitude for red clump stars of $K = -1.61\,\mathrm{mag}$ by \citet{Alves2000}, the assumed intrinsic colour of $H-K_s= 0.07$ of \citet{Nishiyama2006}, a distance to the Galactic Centre of $8\,\mathrm{kpc}$ \citep{Ghez2008} and an approximate extinction for the cluster of $A_{K_s} = 2.35\,\mathrm{mag}$ yields $H_{RC} = 17.05\,\mathrm{mag}$ and $(H-K_{s})_{RC} = 1.79\,\mathrm{mag}$.

Several blue foreground stars with colours $H-K_s \leq 1.3\,\mathrm{mag}$ are seen to the left of the cluster member sequence (Fig.~\ref{f_CMD_member_nonmember_cc}, left panel). These sources are likely disc main sequence stars following the differential rotation of the outer Milky Way rotation curve. With expected velocities of $\sim 200\,\mathrm{km/s}$, they cannot be distinguished from the cluster population on the basis of their proper motion alone.
Furthermore a few very red objects, which could be non-members by comparison with the field CMD, remain in the proper motion sample. In order to remove these contaminants a two-step colour-cut was applied to stars fainter than $H = 14\,\mathrm{mag}$. First the blue foreground and red background stars were removed by keeping only stars with $1.3 \leq H-K_s \leq 2.3\,\mathrm{mag}$. 
In a second step the individual extinction of the remaining stars fainter than $H = 14\,\mathrm{mag}$ was determined from the intersections of the lines of reddening with a $4\;\mathrm{Myr}$ isochrone assuming a distance to the cluster of $8\,\mathrm{kpc}$. The method to derive the individual extinction and the used isochrone are explained in detail in Sect.~\ref{s_mass_derivation}. The isochrone was shifted to an extinction of $A_{K_s} = 2.88\,\mathrm{mag}$, corresponding to the sum of the mean ($\overline{A_{K_s}} = 2.41\,\mathrm{mag}$) and twice the standard deviation ($\sigma_{A_{K_s}} = 0.24\,\mathrm{mag}$) of the individual extinctions of the cluster members remaining after the first colour-cut, and stars redder than the shifted isochrone were also removed from the sample of cluster stars. 

\begin{figure*}
  \resizebox{\hsize}{!}{\includegraphics[angle = 90]{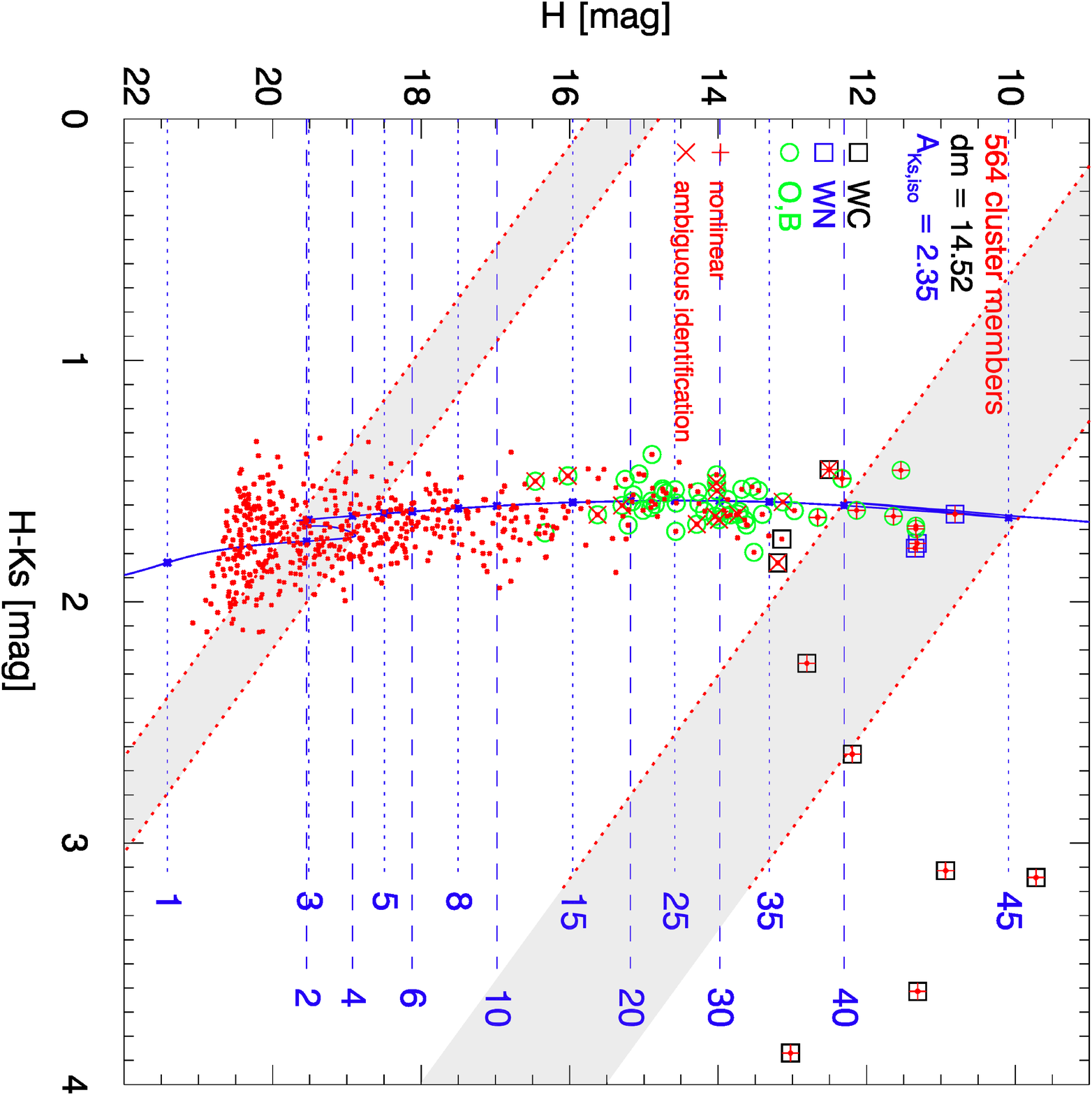}}
  \caption{Colour-magnitude diagram of the final cluster sample. Stars with counterparts in the spectral catalogue of \protect{\citet{Liermann2009}} are flagged with symbols according to their spectral type (box: WR-stars, circle: OB-stars,  stars with ambiguous identification are additionally marked with an X-cross). The horizontal dashed and short-dashed lines mark the initial masses along the isochrone in units of $\mathrm{M_{\sun}}$. The tilted dotted lines show the lines of reddening according to the extinction law by \protect{\citet{Nishiyama2009}} and enframe the two regions in the CMD (shaded in grey in Fig.~\ref{f_CMD_member}, \ref{f_CMD_member_segments} and  \ref{f_CMD_isochrone_comparison}), within which the isochrone has multiple intersections with the line of reddening, and consequently no unique mass can be inferred for a given star.
  }
  \label{f_CMD_member}
\end{figure*}

\begin{figure*}
  \resizebox{\hsize}{!}{\includegraphics[angle = 90]{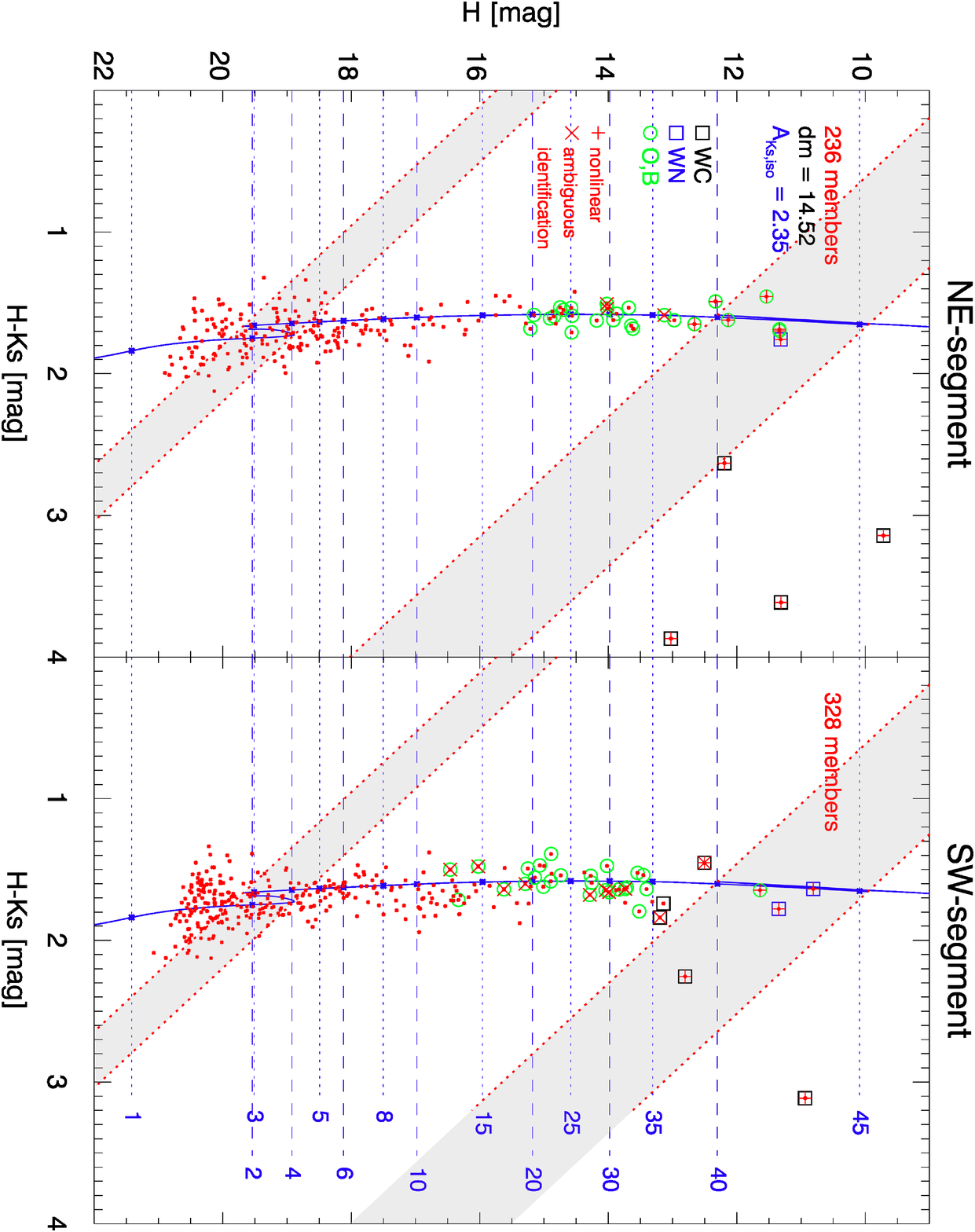}}
  \caption{Colour-magnitude diagram of all stars of the final cluster sample with proper motions residing in the North-East (left panel) or South-West segment (right panel) of the proper motion diagram (Fig.~\ref{f_PMD_member_selection}). See Figs.~\ref{f_CMD_member_nonmember_cc} and \ref{f_CMD_member} for details.}
  \label{f_CMD_member_segments}
\end{figure*}

The designated cluster members and non-members were compared with the K-band spectral catalogue of \citet{Liermann2009} (further abbreviated as LHO catalogue) for a spectral classification of the brighter stars and in order to assess the selection of cluster stars based on their proper motions and colours. Only observed stars with a $K_s$ band magnitude 
brighter than $15.5\,\mathrm{mag}$, which is about $1\,\mathrm{mag}$ fainter than the faintest star in the LHO catalogue, were included in the comparison. 
Eighty-five stars from the spectral catalogue could be assigned to 92 observed stars (69 members, 23 field stars). The ambiguous assignments of 6 stars from the LHO catalogue to 13 observed stars (all members) are caused by the lower spatial resolution of the SINFONI-SPIFFI instrument of $0.250\mathrm{\arcsec}$ for the used $8\mathrm{\arcsec}\times8\mathrm{\arcsec}$ field of view. The spectral classification for the matched stars is indicated in a simplified form by the overplotted symbols in Fig.~\ref{f_CMD_member_nonmember_cc}, Fig.~\ref{f_CMD_member} and Fig.~\ref{f_CMD_member_segments}. Stars with ambiguous assignments are additionally marked with an X-cross. One star (LHO~110) was re-classified in \citet{Liermann2010} from O6-8 I f to WN9h and is treated accordingly in the figures. The numbers and spectral classifications from the LHO catalogue are noted in the source catalogue (Table~\ref{t_source_catalogue}).
Six late-type M,K-giants are still contained within the cluster sample after the colour-cut and are very likely remaining contaminants with motions similar to the cluster members from the Galactic bulge considering the young age of the cluster. These stars and stars rejected by the colour-cut were removed from the final cluster sample and added to the proper motion non-members in the field star CMD (plotted as triangles in the right panel of Fig.~\ref{f_CMD_member_nonmember_cc}).
The one early-type star (O4-7 I f) among the designated field stars is located at the edge of the analysed area of the data and very close to a second star just outside this region. It is therefore unclear if the spectral classification really belongs to this star or its neighbour, hence the star was not added to the final cluster sample. 
For 12 of the 62 stars in the LHO catalogue, which could be assigned to designated cluster members, the sources in our catalogue exceed the $K_s$-band linearity limit by more than $1\,\mathrm{mag}$.  This impedes the repair of the core by \emph{starfinder} and the correct measurement of the position and proper motion. Disregarding these 12 stars, the percentage of contaminating M,K giants, which cannot be discerned from the cluster members based on their proper motion or colour, amounts to $6/(62-12) = 12\%$.

The CMD of the final sample of cluster stars is shown in Fig.~\ref{f_CMD_member} and, separated into the North-East and South-West segment of the proper motion diagram, in Fig.~\ref{f_CMD_member_segments}.
The slight overdensity located at $H-K_s = 1.8\,\mathrm{mag}$, $H = 17\,\mathrm{mag}$ indicates a remaining contamination with red clump stars, which is more pronounced for stars with proper motion in the South-West segment. The CMD for the South-West segment contains 92 stars more than for North-East segment mainly at the faint end of the observed population, which appears slightly broadened. This is expected from the proper motion diagram as the field star population overlaps with the cluster stars in the South-West segment causing  a larger contamination for this segment. The astrometric uncertainty and therefore the scatter in the proper motion diagram increases for fainter magnitudes and therefore the confusion with faint field stars is more severe. The cluster members in the North-East segment therefore constitute the cleanest sample. 

\section{Mass derivation}
\label{s_mass_derivation}

Based on the presence of WC stars, O I stars and a red supergiant within the Quintuplet cluster, \cite{Figer1999} derived an average age of $4\pm1\,\mathrm{Myr}$ assuming a coeval population. More recently the ages of 5 WN stars were determined by comparison of their luminosities and effective temperatures as derived from spectral line fitting with stellar evolution models to be about $2.4-3.6\;\mathrm{Myr}$ pointing to a somewhat younger age of the cluster \citep{Liermann2010}.
To study the influence of the assumed cluster age on the slope of the mass function, three isochrones with ages of $3$, $4$ and $5\,\mathrm{Myr}$ were used to derived the initial stellar masses. The isochrones are a combination of Padova main sequence (MS) isochrones and pre-main sequence (PMS) isochrones derived from Pisa-FRANEC PMS stellar models \citep[see][]{Gennaro2011,Marigo2008,Degl'Innocenti2008}.
As the NACO photometry is calibrated by means of UKIDSS sources (see Sect.~\ref{s_photometric_calibration_and_error_estimation}), the combined  isochrones, for simplicity referred to as $3$, $4$ and $5\,\mathrm{Myr}$ Padova isochrones in the following, were transformed from the 2MASS into the UKIDSS photometric system using the colour equations (6) - (8) from \citet{Hodgkin2009}.
To cover the effect of a different set of stellar models on the derived masses, a $4\,\mathrm{Myr}$ Geneva MS isochrone with enhanced mass loss for high mass stars,  $M>15\,M_{\sun}$, \citep{Lejeune2001} was included in the comparison. The conversion of this isochrone into the UKIDSS filter system encompassed two steps. The isochrone was first transformed from the \citet{Bessell1988} to the 2MASS photometric system using the updated\footnote{Carpenter, J.M., 2003 see \url{http://www.astro.caltech.edu/~jmc/2mass/v3/transformations/}} transformation by \citet{Carpenter2001} and subsequently from the 2MASS to the UKIDSS filter system using the above mentioned conversion. 

\begin{figure*}
  \resizebox{\hsize}{!}{\includegraphics[angle = 90]{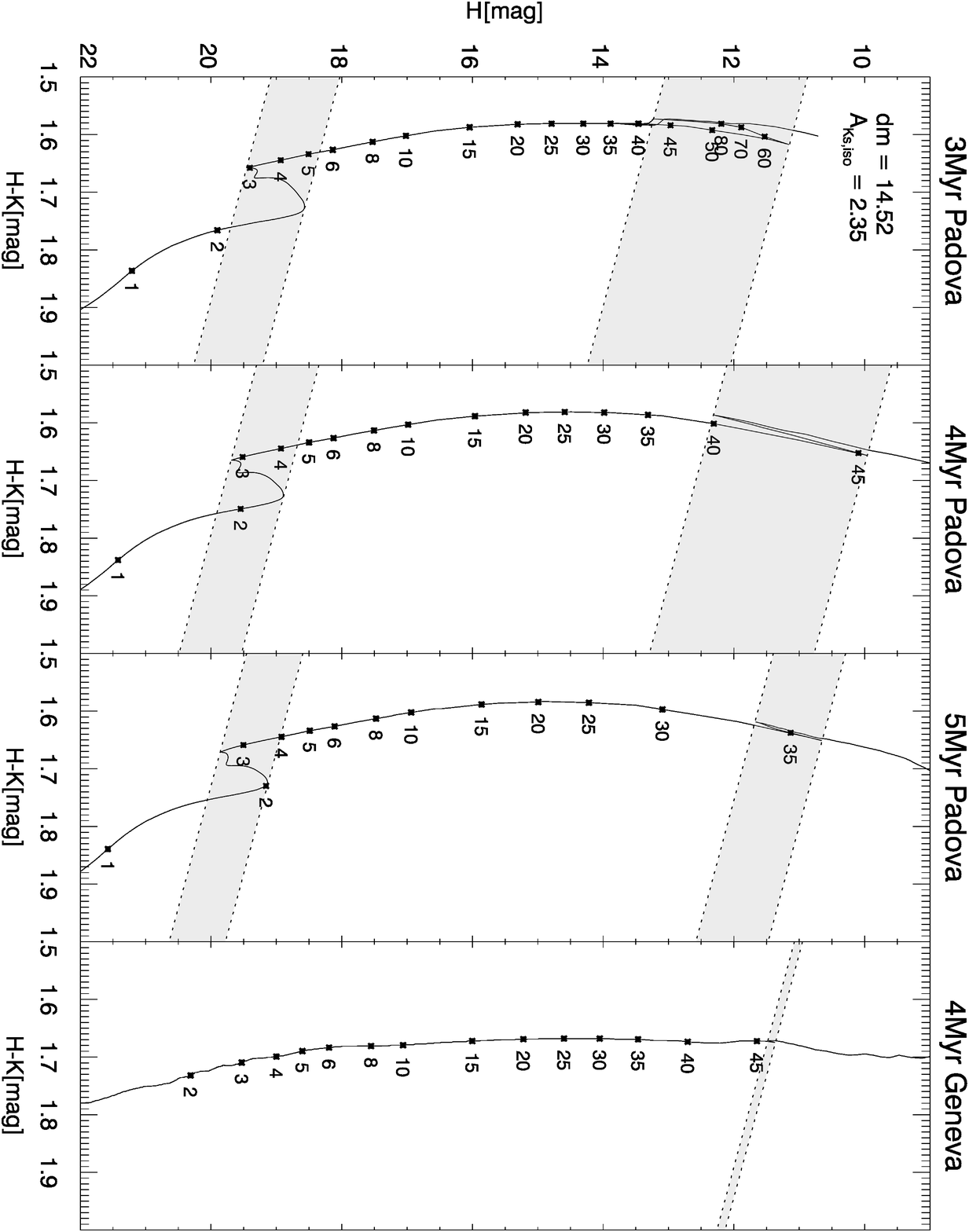}}
  \caption{Comparison of the four isochrones used to determine stellar masses. For all shown isochrones, solar metallicity, a distance to the cluster of  $8\,\mathrm{kpc}$  and a foreground extinction of $A_{K_s} = 2.35\,\mathrm{mag}$ is adopted. As in Fig.~\ref{f_CMD_member} and Fig.~\ref{f_CMD_member_segments}, the dotted lines enframe regions in the CMD with ambiguous mass assignments (shaded in grey) and the initial masses are labelled along the isochrones.}
  \label{f_CMD_isochrone_comparison}
\end{figure*}

For all isochrones,  solar metallicity according to the description of the underlying stellar models\footnote{
solar metallicity for the Geneva isochrone: $\mathrm{X} = 0.68$, $\mathrm{Y} = 0.3$, $Z = 0.020$ \citep{Lejeune2001};\\solar metallicity for the Padova isochrones: $\mathrm{X} = 0.708$, $\mathrm{Y} = 0.273$, $Z = 0.019$ for $M<7\,M_{\sun}$ \citep{Marigo2008,Girardi2000}, $\mathrm{X} = 0.7$, $\mathrm{Y} = 0.28$, $Z = 0.020$ for $M>7\,M_{\sun}$ \citep{Bertelli1994,Bressan1993}} was assumed, and a distance to the Galactic centre of $8.0\,\mathrm{kpc}$ \citep{Ghez2008} was applied as the distance to the Quintuplet cluster. 
The four isochrones shown in Fig.~\ref{f_CMD_isochrone_comparison} were reddened by a foreground extinction of $A_{K_s} = 2.35\,\mathrm{mag}$ using the extinction law of \citet{Nishiyama2009} ($A_H:A_{K_s} = 1.73:1$) to match the observed MS of the cluster members. This extinction law is one of the most recent determinations of the extinction in the near-infrared along the line of sight towards the Galactic centre and consistent with other current findings, e.g., by \citet{Straivzys2008} or \citet{Schodel2010}.

The individual mass and extinction of each star in the final cluster sample was determined from the intersection of the line of reddening through the star with the respective isochrone in the CMD. Due to the local maximum of the PMS at the low-mass end as well as the extended loop at the transition from the end of the hydrogen core burning to the contraction phase at the high-mass end, the de-reddening path of a star may have several intersections with the isochrone, thus leading to an ambiguous mass assignment (the affected areas in the colour-magnitude plane are shaded in grey in Figs.~\ref{f_CMD_member}, \ref{f_CMD_member_segments} and  \ref{f_CMD_isochrone_comparison}).
For these stars, the masses at each intersection were averaged. 
The post-MS phase after the exhaustion of hydrogen in the stellar core is very rapid (a few $10^3\,\mathrm{yr}$ according to the stellar models) and apparent in the isochrones as the branch with increasing $H$-band brightnesses, re-rising after the decline connected to the contraction phase. Due to its short duration, which causes the Hertzsprung gap in the Hertzsprung-Russel diagrams of stellar clusters, only the two intersection points with the upper part of the MS and with the subsequent falling branch of the isochrone were averaged. 
Two O~stars from the LHO catalogue have no intersection with the Geneva isochrone on the MS or the falling branch, therefore an initial mass of $47.3\,\mathrm{M_{\sun}}$, which is the maximum mass along this isochrone used for the mass determination (see Table~\ref{t_mass_determination}),  was assigned to them. 

11 Wolf-Rayet stars out of the 21 observed in the Quintuplet Cluster are contained within our sample of cluster members. The masses for these stars  could not be determined from the isochrones but the mass range of Wolf-Rayet stars was inferred from the underlying stellar models by \citet{Bressan1993} for the Padova isochrones and by \citet{Meynet1994} and \citet{Schaller1992} for the Geneva isochrone (see Table~\ref{t_mass_determination}).
 
\begin{figure*}
  \resizebox{\hsize}{!}{\includegraphics[angle = 90]{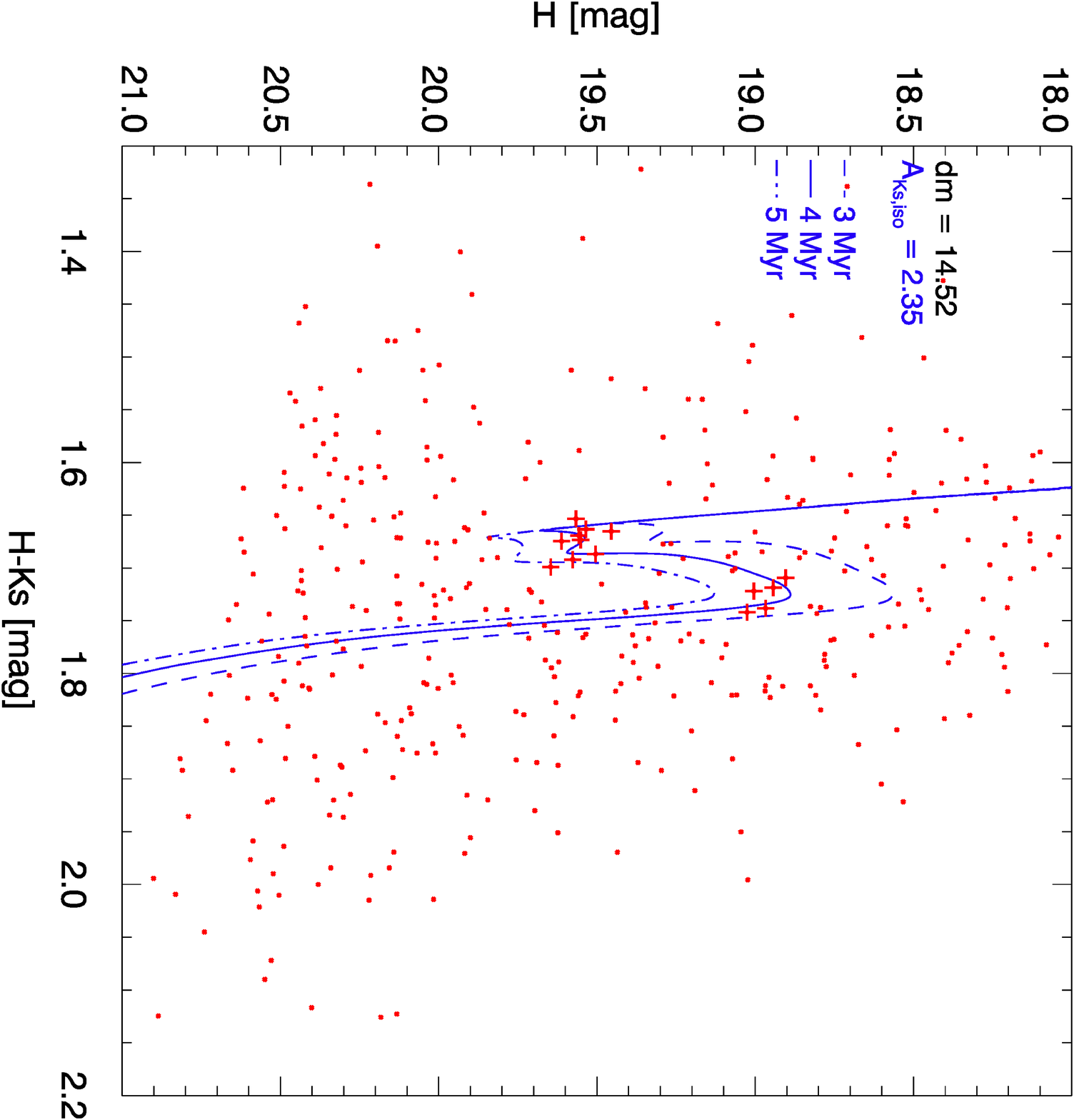}}
  \caption{Detail of the transition region between PMS and MS in the member CMD (Fig.~\ref{f_CMD_member}). The three Padova isochrones (with PMS parts derived from Pisa-FRANEC PMS stellar models), shifted to an extinction of $A_{K_s} = 2.35\,\mathrm{mag}$ to fit the MS, are shown as well. The stars comprising the overdensities referred to in Sect.~\ref{s_colour_magnitude_diagrams} are marked with crosses. The clump of stars at the MS turn-on point is particularly striking, and matches excellently the predicted MS turn-on point of the $4\,\mathrm{Myr}$ Padova isochrone.}
  \label{f_CMD_member_PMS_MS}
\end{figure*}

Figure~\ref{f_CMD_member_PMS_MS} shows an enlarged detail of the transition region from the PMS to the MS  in the member CMD. The three Padova isochrones shown are all shifted to an extinction of $A_{K_s} = 2.35\,\mathrm{mag}$ in order to fit the cluster MS.
Although stars scatter across the whole transition region, there is an indication of a slight accumulation of stars at the turn-over of the PMS for a cluster age of $4\,\mathrm{Myr}$. At this stage of evolution the \element{CN} cycle supports most of the stellar radiative losses, but the \element{NO} cycle is still not in complete equilibrium. After sufficient \element{N} is produced, the star undergoes a rapid contraction and reaches the MS after the \element{CN}-\element{NO} cycle is in complete equilibrium. The MS-turn-on point for this cluster age is marked by a second overdensity of stars (see Fig.~\ref{f_CMD_member_PMS_MS}).
In contrast, the turn-over of the PMS and MS-turn-on point of the $5\,\mathrm{Myr}$ isochrone as well as the MS-turn-on for the $3\,\mathrm{Myr}$ isochrone are located in a region almost devoid of stars.
Irrespective of the definition of the "true" age of the cluster, we therefore consider the $4\,\mathrm{Myr}$ Pisa-FRANEC/Padova isochrone as the best fitting isochrone within the investigated set of models.

\begin{table*}
\caption{Summary of isochrone properties relevant for the mass derivation.}
\label{t_mass_determination}
\begin{tabular}{clrcccc}
\hline\hline
Isochrone name& Description & Wolf-Rayet mass  & \multicolumn{4}{c}{stars with ambiguous mass assignments}\\
              &             &                  & \multicolumn{2}{c}{PMS $\rightarrow$ MS} & \multicolumn{2}{c}{MS  $\rightarrow$  post-MS}\\
	      &		    &                  & {mass range} & No. of stars & {mass range} & No. of stars\\
	      &             & ($\mathrm{M_{\sun}}$) & ($\mathrm{M_{\sun}}$) & &  ($\mathrm{M_{\sun}}$)\\
$3\,\mathrm{Myr}$ Padova  & MS\tablefootmark{a} + PMS\tablefootmark{b} & $85-100$ & $2.1-5.5$  & 145 & $42.3-84.4$ & 11\\
$4\,\mathrm{Myr}$ Padova  & MS\tablefootmark{a} + PMS\tablefootmark{b} & $51-65$  & $1.9-4.6$  & 145 & $39.9-50.8$  &  5\\
$5\,\mathrm{Myr}$ Padova  & MS\tablefootmark{a} + PMS\tablefootmark{b} & $37-40$  & $1.8-4.0$  & 149 & $33.5-35.4$  &  4\\
$4\,\mathrm{Myr}$ Geneva  & MS\tablefootmark{c}                        & $48-60$  &              &   & $45.8-47.3$  &  0\\
\end{tabular}

\tablefoot{
  \tablefoottext{a}{Padova isochrone with solar metallicity for $m > 4\,\mathrm{M_{\sun}}$ \protect{\citep{Marigo2008}}.}
  \tablefoottext{b}{Pre-main sequence parts of the isochrones ($m \leq 4\,\mathrm{M_{\sun}}$) are derived from Pisa-FRANEC PMS stellar models (\protect{\citet{Degl'Innocenti2008}}, see \protect{\citet{Gennaro2011}} for the combination with the Padova isochrones).}
  \tablefoottext{c}{Geneva isochrone with solar metallicity and enhanced mass loss for high mass stars, $M>15\,M_{\sun}$, \protect{\citep{Lejeune2001}}.}
}

\end{table*}

\renewcommand\tabcolsep{3pt}
\begin{sidewaystable}
\centering
\vspace{100mm}
\caption{Catalogue of stellar sources with measured proper motions and colours in the Quintuplet cluster.}
\label{t_source_catalogue}
\begin{tabular}{lcccccccccccccccccc}
\hline\hline
No. & $\Delta \mathrm{R.A.}$\tablefootmark{a} & $\Delta \mathrm{Decl.}$\tablefootmark{a} & $K_s$ &$\sigma_{K_s}$ &  $H$ &$\sigma_{H}$ & compl.\tablefootmark{b} & $\mu_{\alpha \cos{(\delta)}}$& $\mu_{\delta}$ & $\sigma_{\mu}$\tablefootmark{c} &segment\tablefootmark{d}& LHO No.\tablefootmark{e}& Type \tablefootmark{e}&member\tablefootmark{f}& $\mathrm{m_{Pad,3\,Myr}}$\tablefootmark{g} & $\mathrm{m_{Pad,4\,Myr}}$\tablefootmark{g} & $\mathrm{m_{Pad,5\,Myr}}$\tablefootmark{g} & $\mathrm{m_{Gen,4\,Myr}}$\tablefootmark{g}\\
  & ($\arcsec$) & ($\arcsec$) & (mag) & (mag) &  (mag) & (mag) &  & $\mathrm{(mas/yr)}$ &$\mathrm{(mas/yr)}$ & $\mathrm{(mas/yr)}$ & & & & & $\mathrm{(M_{\sun})}$ & $\mathrm{(M_{\sun})}$  & $\mathrm{(M_{\sun)}}$ & $\mathrm{(M_{\sun})}$\\
1&-7.58&3.99&7.70&0.12&11.31&0.05&0.99&-2.30&0.91&1.61&NE&75&WC9?d&y&85-100&51-65&37-40&48-60\\
2&-5.16&9.25&9.16&0.12&13.03&0.03&1.00&-2.13&-0.87&0.57&NE&102&WC9?d&y&85-100&51-65&37-40&48-60\\
3&-0.00&0.00&6.58&0.10&9.72&0.05&1.00&-2.09&0.25&1.61&NE&42&WC9d + OB&y&85-100&51-65&37-40&48-60\\
4&1.09&6.45&7.82&0.08&10.94&0.02&1.00&0.93&0.54&1.61&SW&84&WC9d&y&85-100&51-65&37-40&48-60\\
5&5.35&3.70&9.17&0.08&10.81&0.02&1.00&-0.59&-1.21&0.38&SW&71&WN9&y&85-100&51-65&37-40&48-60\\
6&16.21&3.04&9.56&0.19&11.32&0.13&1.00&-1.21&0.52&1.40&NE&67&WN9&y&85-100&51-65&37-40&48-60\\
7&9.04&6.26&9.56&0.05&12.19&0.02&1.00&-0.37&1.13&0.37&NE&79&WC9d&y&85-100&51-65&37-40&48-60\\
8&5.06&11.40&9.57&0.07&11.35&0.06&1.00&0.60&-0.30&0.81&SW&110&WN9h&y&85-100&51-65&37-40&48-60\\
9&6.08&9.30&9.65&0.08&11.34&0.03&1.00&-0.46&0.29&0.45&NE&100&O6-8 I f e&y&56.38&46.05&34.53&47.28\\
10&5.65&8.24&9.64&0.07&11.34&0.04&1.00&0.50&0.50&0.39&NE&96&O6-8 I f e&y&56.30&46.00&34.50&47.28\\
\end{tabular}
\tablefoot{This table is available in its entirety in a machine-readable form at the CDS via anonymous ftp to \url{cdsarc.u-strasbg.fr} or via \url{http://cdsweb.u-strasbg.fr/cgi-bin/}. A portion is shown here for guidance regarding its form and content.\\
\tablefoottext{a}{Positional offset in right ascension and declination relative to the AO guide star Q2 (R.A. = 17:46:14.690, Dec. = -28:49:40.71 [J2000])}.
\tablefoottext{b}{The combined completeness for each star is the product of its completeness in the $K_s$-band datasets of both epochs (2003 and 2008) and in the H-band dataset from 2003.}
\tablefoottext{c}{The uncertainty of the proper motion includes the astrometric uncertainties in both epochs and the rms of the geometric transformation in x- and y- direction.}
\tablefoottext{d}{Indicates the segment in the proper motion diagram in which the star resides.}
\tablefoottext{e}{Numbers and spectral identifications from the spectral catalogue by \protect{\citet{Liermann2009}}. Sources in the LHO catalogue which have more than one counterpart in this table are marked with an asterisk.}
\tablefoottext{f}{Cluster membership: cluster members are indicated by "y", field stars according to their measured proper motions are indicated by "n".  Proper motion members rejected based on their spectral type (M,K giants) or their colour are marked with 'n st' or 'n cc', respectively.}
\tablefoottext{g}{Initial masses as determined from the $3$, $4$, $5\,\mathrm{Myr}$ Padova isochrones (with PMS part derived from Pisa-FRANEC PMS stellar models, see \protect{\citealt{Gennaro2011}}) and the $4\,\mathrm{Myr}$ Geneva isochrone. All isochrones assume solar metallicity. For Wolf-Rayet stars the stated mass ranges are inferred from the underlying stellar models of the respective isochrone.}
}
\end{sidewaystable}
\renewcommand\tabcolsep{6pt}

\section{Mass functions}
\label{s_mass_functions}

In order to avoid potential biases introduced by bins with a very small number of objects or large differences in the number of stars between the low- and high-mass bins, we adopted the method proposed by \citet{Apellaniz2005}. Here, the widths of the different bins are adjusted such that each bin houses approximately the same number of stars (Method A).
If the number of stars in the sample did not split up evenly for the chosen number of bins, the bins to contain one additional star from the remaining stars were chosen randomly. The stars were then sorted according to their masses and distributed among the bins. For each isochrone the mass function (MF) and slope were determined for dividing the cluster sample into $4$, $8$, $12$, $16$ and $20$ bins. 
The boundary between two adjacent bins was set to the mean of the most/least massive star in the respective bins.  
The minimum mass used for each mass function was set to the lowest mass of a star with a unique mass assignment for the respective isochrone, i.e. lying above the ambiguity region caused by the PMS/MS transition. Stars with ambiguous mass assignments at the upper end of the MS  were kept for the mass function, as due to their small number they all contribute to the uppermost bin in the mass function. The upper mass limit or uppermost bin boundary $m_{\mathrm{up}}$ was calculated from the data to be (see \citet{Apellaniz2005})
\begin{equation}
      \label{e_upper_mass_limit}
      m_{\mathrm{up}} = m_n + 0.5\left(m_n-m_{n-1}\right)\,,
\end{equation}
with $n$ being the total number of stars.
The number of stars in each bin $n_i$ was normalized by the respective bin width $\Delta m_i$.
The logarithm of the normalized number of stars per bin as a function of the logarithm of the medium mass of each bin was fitted with a straight line using the IDL routine \emph{LINFIT}, which performs a $\chi^2$ minimisation.

The uncertainty of the number of stars per bin $\Delta n_i$ is derived by \citet{Apellaniz2005} from the standard error of a binomial distribution $(n p_i \left(1-p_i\right))^{1/2}$, where the unknown true probability for a star to  reside in the $i^{\mathrm{th}}$ bin $p_i$ is approximated by the measured value $n_i/n$: 
\begin{equation}
  \Delta{n_i} = \sqrt{\frac{n_i \left(n-n_i\right)}{n}}\,.
\end{equation}
Note that this uncertainty differs from the Poisson error $\sqrt{n_i}$, which is usually applied to binned data. 
For the linear fit each bin was weighted by its statistical weight $w_i = 1 / {\Delta n_i}^2.$
The statistical weight $w_i$ assigned to the logarithm of the normalized number of stars per bin ($\log_{10}(n_i/{\Delta m_i})$) follows from error propagation of $\Delta{n_i}$ in the logarithmic plane (see equation (7) in \citealt{Apellaniz2005}):
\begin{equation}
    \label{e_weight}
      w_i = \frac{n\,n_i\,2\ln{10}}{n -n_i}\,.
\end{equation}
It is basically the same for every bin as the number of stars per bin varies by a maximum of one. 

Besides the binning method just described the mass function was also determined using an equal logarithmic width for each bin (Method B), which is still the most common binning method for deriving mass function slopes (see \citet{Apellaniz2005} for a discussion of the biases of this method). The lower and upper mass limits were determined in exactly the same way as above and the logarithmic bin widths were set by dividing the so defined mass range into $4$, $8$, $12$, $16$ and $20$ bins. The weights applied to each bin were again calculated with equation~(\ref{e_weight}) and are decreasing going to higher masses due to the lower number of stars contained in the high mass bins. In order to study the influence of the weights on the slope for this binning method, the slope was derived from a linear fit to the MF with and without weighting.
\newline

The reported slopes of the mass function $\alpha$ refer to a power-law distribution in linear units ($dn/dm \propto m^{\alpha}$) with the standard Salpeter slope being $\alpha = -2.35$ in this notation \citep{Salpeter1955}. 
If not mentioned otherwise, the mass function and its slope were determined using all cluster members from both the North-East and South-West segment (see Fig.~\ref{f_CMD_member_segments}) and distributing the stars into bins with (almost) constant number of stars (Method A). 
All shown linear fits to the respective mass functions were derived from the completeness corrected mass function using for each star its individual completeness correction (see Sect.~\ref{s_completeness}).

The minimum mass of a star with unique mass assignment was $5.5$, $4.6$ and $4.0\,\mathrm{M_{\sun}}$ for the $3$, $4$ and $5\,\mathrm{Myr}$ Padova isochrone, respectively. The minimum mass for the $4\,\mathrm{Myr}$ Geneva isochrone was set to $4.5\,\mathrm{M_{\sun}}$ in order to use exactly the same stars as for the Padova isochrone of the same age.
As mentioned in Sect.~\ref{s_mass_derivation}, it was not possible to infer the individual masses of the WR stars from the isochrones. Therefore, a constant mass within the mass ranges of the Wolf-Rayet stars deduced from the stellar models (see Table~\ref{t_mass_determination}) was assigned to each identified Wolf-Rayet star in the cluster sample in dependence of the assumed cluster age. The uppermost bin boundary, calculated with equation~(\ref{e_upper_mass_limit}), is then identical to the assigned WR mass. 
The chosen WR mass has a significant impact on the derived slopes due to the fairly large mass range of the Wolf-Rayet stars for cluster ages of $3$ and $4\,\mathrm{Myr}$. A larger assigned WR mass biases the mass function to a steeper slope due to the normalization of $n_i$ by the binwidth. The maximum difference between the slopes using the minimum and maximum WR masses for each of the isochrones was $0.23$, which is about twice the typical formal fitting error of the slope.
To avoid the described bias, the Wolf-Rayet stars were not included in the mass function. After the exclusion of the Wolf-Rayet stars, the uppermost bin boundary is determined by the two most massive stars in the respective sample (see equation~\ref{e_upper_mass_limit}).

In order to quantify the effect of the random selection of bins to contain one additional star from the remainder of the division of the total number of stars by the number of bins (Method A), the distribution process and the fit to the resulting MF was repeated $1000$ times. The reported slopes for this binning method are the mean slope of all these repetitions. The maximum difference between the slopes of the same MF due to different random distributions of the surplus stars was $0.02$, which is very small compared to the formal fitting errors.

The slope of the mass function of each isochrone was determined using $4$, $8$, $12$, $16$ and $20$ bins. Using only $4$ bins results in slopes being systematically shallower than for the other numbers of bins by up to  $0.12$.
The choice of $20$ bins introduces a bias in the case of the $3\,\mathrm{Myr}$ isochrone. Due to the large number of massive stars with averaged masses for this isochrone (see Table~\ref{t_mass_determination}), these stars fill up the uppermost bin completely.
The mass range of stars with averaged masses is compressed, which in turn leads to a decreased binwidth of the last bin. As the number of stars is normalized by the bin width, the normalized number of stars in the last bin is increased leading to a flatter slope.
The most reliable mass function slopes are therefore obtained using $8$, $12$ or $16$ bins. The maximum difference of the obtained slopes for a given isochrone between these three bin numbers was $0.03$. 
Given this negligible influence of the bin number, all results presented in the following are determined using $12$ bins (see also Table~\ref{t_mass_function}).

\begin{figure*}
  \resizebox{\hsize}{!}{\includegraphics{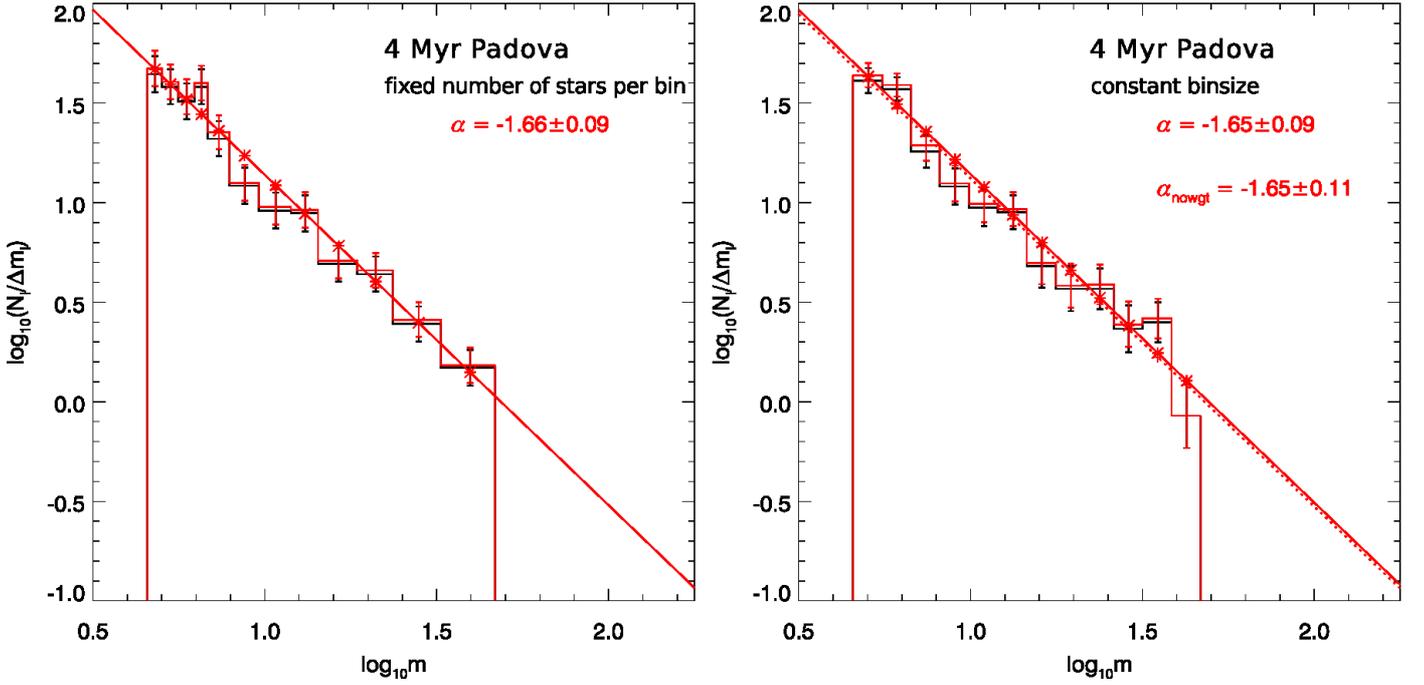}}
  \caption{Comparison of the mass function  and derived slopes for different methods of binning the data and performing the linear fit. Only  the fit and derived slope for the completeness corrected mass function are shown.
  Left panel: Mass function of the Quintuplet Cluster derived from the colour-magnitude diagram in Fig.~\ref{f_CMD_member} using a MS-Padova isochrone and a cluster age of $4\,\mathrm{Myr}$ to transform magnitudes into stellar masses. Only stars with $m > 4.6\,\mathrm{M_{\sun}}$, i.e. stars above the ambiguity region in the CMD due to the PMS, are used. Wolf-Rayet stars are not included in the mass function, as the large uncertainty of their mass might bias the derived slopes. The bin sizes are adjusted such that each of the 12~ bins holds approximately the same number of stars.
  Right Panel: Mass function of the same data but distributing the stars into 12 bins of a uniform logarithmic width of $0.084\,\mathrm{dex}$ adopting the same lower and upper mass limits as in the left panel. The solid line shows the weighted linear fit, the dotted line the unweighted fit. The slopes of both methods are in excellent agreement.
}
  \label{f_MF_method}
\end{figure*}

Figure~\ref{f_MF_method} shows the comparison of the mass functions and the derived slopes for distributing the data in bins with (almost) constant number of stars (Method A, left panel) and for using bins of equal logarithmic width of $\Delta \log_{10}{m} = 0.084\,\mathrm{dex}$ (Method B, right panel) for the $4\,\mathrm{Myr}$ Padova isochrone. In the shown example and in general the three slopes of the weighted fit to the mass function derived with a uniform number of stars per bin and the weighted and unweighted fit to the mass function with an equal logarithmic bin widths agree well within the errors, if the full sample of stars is fitted.
The use of a constant logarithmic bin size with applied weights following the prescription of \citet{Apellaniz2005} generates consistent results compared to the use of bins with variable widths and equal numbers of stars also if only stars from the North-East or South-West segment of the proper motion diagram are included in the mass function.  
In contrast the unweighted fit responds much more sensitively to fluctuations of the number of stars in the higher mass bins, especially if the last  bin included in the fit is depleted, which is the case for the South-West segment.
For the remainder of this paper only the results determined from mass functions with an equal number of star per bin are considered.

\begin{figure*}
  \resizebox{\hsize}{!}{\includegraphics{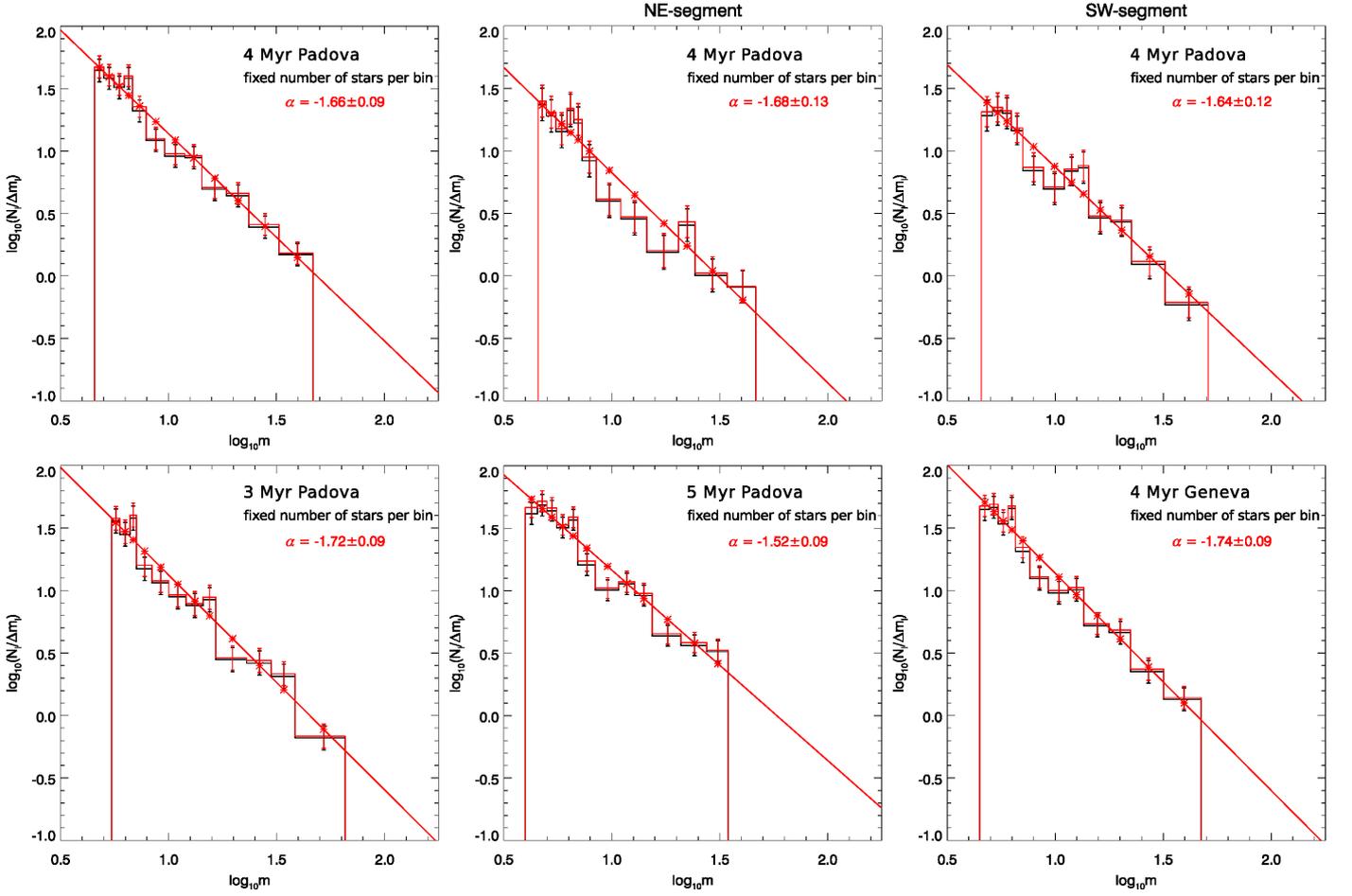}}
  \caption{Upper panels: Mass functions for the $4\,\mathrm{Myr}$ MS-Padova isochrone using all stars (left), or only stars located in the North-East- (middle), or South-West-segment (right) of the proper motion diagram (Fig.~\ref{f_PMD_member_selection})
  Lower panels: Resulting mass function if stellar masses are derived from a $3$ or $5\,\mathrm{Myr}$ Padova MS-isochrone (left and middle) or $4\,\mathrm{Myr}$ Geneva isochrone (right).
  For all shown mass functions the stars were distributed in 12 bins of variable width with equal numbers of stars per bin. As in Fig.~\ref{f_MF_method}, the Wolf-Rayet stars were removed and only stars above the ambiguity region in the CMD due to the PMS are included. The resulting minimum masses are $5.5$, $4.6$, $4.0$, and $4.5\,\mathrm{M_{\sun}}$, for the $3$, $4$, $5\,\mathrm{Myr}$ Padova and the $4\,\mathrm{Myr}$ Geneva isochrone, respectively.}
  \label{f_MF_comp}
\end{figure*}

The first row in Fig.~\ref{f_MF_comp} shows the mass function for the $4\,\mathrm{Myr}$ Padova isochrone for all cluster members (left panel), for stars in the North-East-segment (centre panel), and for stars in the South-West segment (right panel). All three slopes agree well, although a lack of stars in the mass function of the North-East segment in the mass range of $11-22\,\mathrm{M_{\sun}}$ compared to the South-West segment is evident. This surplus of stars for the South-West segment is also apparent in the CMD (Fig.~\ref{f_CMD_member_segments}) and could indicate a remaining contamination with red clump stars. Further contaminations suggested by the difference in the number of cluster members in the North-East and South-West segment are likely removed by only using stars with intermediate brightness and masses above $4.6\,\mathrm{M_{\sun}}$.

For the Padova isochrones the slope of the mass function decreases with increasing age going from $3$ to $5\,\mathrm{Myr}$  from $\left(-1.72 \pm 0.09\right)$ to $\left(-1.52 \pm 0.09\right)$. As can be seen in Fig.~\ref{f_CMD_isochrone_comparison}, the range of initial masses along the upper part of the MS starting at about $20\,\mathrm{M_{\sun}}$ strongly decreases with age. This causes the same number of brighter stars in the CMD being squeezed into a smaller mass interval for the older ages, which therefore results in a flattening of the slopes with increasing age.
The initial masses derived using the $4\,\mathrm{Myr}$ Padova isochrone are $0\%-7\%$ larger for stars with $m < 37\,\mathrm{M_{\sun}}$ than the initial masses determined with the $4\,\mathrm{Myr}$ Geneva isochrone. For stars with even higher masses the Geneva isochrone yields slightly larger masses. This varying difference in the deduced masses causes the slope derived from the Geneva isochrone to be steeper by $0.08$ in comparison with the Padova isochrone of the same age. Nonetheless, the two derived slopes agree within the fitting uncertainties.
\newline

All slopes derived binning the data into 12 bins containing approximately the same number of stars per bin are summarized in Table~\ref{t_mass_function}. The slopes are all internally consistent: 1.) For each isochrone the slope derived for the South-West segment is flatter than the slope for the North-East segment, and the slope of the full sample is very close to the average of the slopes of both segments. 2.) Independent of the sample (NE + SW, NE, SW), the slope of the mass function decreases with the assumed cluster age for the Padova isochrones, and the use of the $4\,\mathrm{Myr}$ Geneva isochrone results in a steeper slope than for the $4\,\mathrm{Myr}$ Padova isochrone.
For the North-East and South-West segment all slopes agree within the formal fitting uncertainties irrespective of the isochrone. 
For the full sample the error margins of the slope derived for the $5\,\mathrm{Myr}$ Padova isochrone have just no overlap with the error margins of the slopes derived for the $3\,\mathrm{Myr}$ Padova, the $4\,\mathrm{Myr}$ Padova and the $4\,\mathrm{Myr}$ Geneva isochrone.
The average value of the slopes of all considered isochrones, using the full sample of cluster members, therefore provides a robust estimate for the mass function slope of the Quintuplet cluster. The average slope is $-1.66$, which is the same value as the slope of the best-fitting $4\,\mathrm{Myr}$ Padova isochrone. The maximum difference  between this average and the four regarded slopes is $0.14$, which provides a conservative estimate of the uncertainty of the average slope.

\begin{table*}
\caption{Overview of derived slopes of the mass function binning the data into $12$ bins containing approximately the same number of stars.}
\label{t_mass_function}

\begin{tabular}{llccccccc}
\hline\hline
Isochrone name& segment\tablefootmark{a} & No. of stars  & $n_i$\tablefootmark{b} & $m_{\mathrm{min}}$ &  $m_{\mathrm{max}}$ & $\alpha$ \tablefootmark{c}& $\Delta\alpha_{\mathrm{fit}}$\tablefootmark{d} & $\Delta\alpha_{\mathrm{binning}}$\tablefootmark{e}\\
              &         &               & 	&($\mathrm{M_{\sun}}$) & ($\mathrm{M_{\sun}}$)    &          &    &\\
$3\,\mathrm{Myr}$ Padova  & NE + SW & 219  & 18-19 & 5.5 & 65.8 & -1.72 & 0.09 & 0.02\\
$4\,\mathrm{Myr}$ Padova  & NE + SW & 257  & 21-22 & 4.6 & 46.7 & -1.66 & 0.09 & 0.01\\
$5\,\mathrm{Myr}$ Padova  & NE + SW & 283  & 24-25 & 4.0 & 34.5 & -1.52 & 0.09 & 0.02\\
$4\,\mathrm{Myr}$ Geneva  & NE + SW & 257  & 21-22 & 4.5 & 47.3 & -1.74 & 0.09 & 0.01\\
\hline
$3\,\mathrm{Myr}$ Padova  & NE      & 100  & 8-9   & 5.5 & 65.8 & -1.75 & 0.13 & 0.06\\
$4\,\mathrm{Myr}$ Padova  & NE      & 119  & 10-11 & 4.6 & 46.4 & -1.68 & 0.13 & 0.01\\
$5\,\mathrm{Myr}$ Padova  & NE      & 131  & 11-12 & 4.0 & 34.5 & -1.60 & 0.13 & 0.01\\
$4\,\mathrm{Myr}$ Geneva  & NE      & 119  & 10-11 & 4.5 & 47.3 & -1.76 & 0.13 & 0.01\\
\hline
$3\,\mathrm{Myr}$ Padova  & SW      & 119  & 10-11 & 5.5 & 65.6 & -1.71 & 0.13 & 0.01\\
$4\,\mathrm{Myr}$ Padova  & SW      & 138  & 11-12 & 4.6 & 51.0 & -1.64 & 0.12 & 0.02\\
$5\,\mathrm{Myr}$ Padova  & SW      & 152  & 12-13 & 4.0 & 36.3 & -1.50 & 0.12 & 0.02\\
$4\,\mathrm{Myr}$ Geneva  & SW      & 138  & 11-12 & 4.5 & 48.1 & -1.68 & 0.12 & 0.02\\
\end{tabular}
\tablefoot{
  \tablefoottext{a}{Segment in the proper motion diagram (Fig.~\ref{f_PMD_member_selection}).}
  \tablefoottext{b}{Number of stars per bin.}
  \tablefoottext{c}{Average of the slopes  derived for 1000 realizations of randomly distributing the remainder of the division of the number of stars by the number of bins into the bins by increasing the number of stars in the selected bin by one.}
  \tablefoottext{d}{Formal uncertainty of the linear fit.}
  \tablefoottext{e}{Maximum difference between the slopes due to the random distribution of the surplus stars.}
}
\end{table*}

Our best value of the slope of the present-day mass function of the Quintuplet cluster for stars within a radius of $0.5\;\mathrm{pc}$ from the cluster centre, an inital mass of $m_{init} > 5\,\mathrm{M_{\sun}}$ and excluding spectroscopically identified Wolf-Rayet stars, is $\alpha = -1.66\pm0.14$. 
It should be noted that we determined the slope using the \emph{initial} masses as inferred from the isochrones. Furthermore, binarity is not accounted for, as we cannot observe a binary sequence in the CMD of the Quintuplet cluster. Therefore, the reported slopes refer to the system mass function. \citet{Weidner2009} performed a numerical study to determine the influence of unresolved multiple systems on the initial mass function. Assuming 100\% of the stars being part of multiple systems and using three different pairing methods they find that the difference of the slopes of the single star and the observed system IMF for the high mass stars ($m > 2\,\mathrm{M_{\sun}}$) is in general smaller than the usual error bars of observational slopes. In general, the system IMF tends to be steeper by about $0.1$ than the single star IMF.
\citet{Rio2009} derive a maximum difference between the single star and the system IMF of $0.2$ for the mass range of $1 < M_{\sun} < 20$ by using random pairing and varying the binary fraction between 0 and 1. 
The flat MF slope of $\alpha = -1.66$ observed in the Quintuplet cluster within a radius of $0.5\,\mathrm{pc}$ compared to the canonical slope of $-2.3$ is therefore not mimicked by non-resolved multiple systems. 

The total mass of stars in the final cluster sample amounts to $4355\,\mathrm{M_{\sun}}$ adopting the initial masses derived from the $4\,\mathrm{Myr}$ Padova isochrone and an average mass of $58\,\mathrm{M_{\sun}}$ for each of the 11 WR stars. Extrapolating the MF down to a minimum mass of $0.5\,\mathrm{M_{\sun}}$ results in a total mass of the Quintuplet cluster within a radius of $0.5\,\mathrm{pc}$ of $5895\,\mathrm{M_{\sun}}$.

\section{Discussion}
\label{s_discussion}

All derived slopes of the mass function in the central part of the Quintuplet cluster above a mass of $5\,\mathrm{M_{\sun}}$ are systematically flatter than the canonical slope of the inital mass function of $\alpha = -2.3\pm0.7$ for the same mass regime \citep{Kroupa2001}, albeit still marginally contained within its large $99\%$ confidence limits. This indicates that the cluster  within a radius of $0.5\,\mathrm{pc}$ is depleted of lower mass stars.

This result is not unexpected with respect to findings in other Galactic young massive clusters (see Table~\ref{t_slope_comparison}), which expose signs of mass segregation by steepening slopes of their mass function for larger distances to the cluster centre.
The mass function of the young cluster NGC~3603 (age $1-2.5\,\mathrm{Myr}$) exposes a gradually steeper slope for larger annuli from $\alpha = -1.31$ within  $R < 0.15\,\mathrm{pc}$ to $-1.75$ for $0.4<R<0.9\,\mathrm{pc}$ \citep{Harayama2008}. Up to a maximum observed distance of $3.3\,\mathrm{pc}$ from the assumed cluster centre the slope remains almost constant ranging from $-1.80$ to $-1.86$. The global slope of $-1.74$ for $0.4 < m < 20\,\mathrm{M_{\sun}}$  is well below the canonical IMF slope of $-2.3$, suggesting  a top-heavy IMF for this cluster. 
Westerlund~1, with an age of about $3$ to $5\,\mathrm{Myr}$, exposes a flattened MF with $\alpha = -1.6$ for stars in the mass range of $3.4 < m < 27\,\mathrm{M_{\sun}}$ within $R < 0.75\,\mathrm{pc}$, which successively steepens at larger radii up to $\alpha = -2.7$  for $R > 2.1\,\mathrm{pc}$ \citep{Brandner2008}. These general findings were confirmed in a follow-up paper by \citet{Gennaro2011}, which drops the assumption of radial symmetry for the cluster and determines the mass function in a two-dimensional approach. Their global mass function slope is with $\alpha = -2.55^{+0.20}_{-0.08}$ even steeper than the canonical slope. 
The Arches cluster (age $\sim2.5\,\mathrm{Myr}$), is located at a projected distance to the Galactic centre of $26\,\mathrm{pc}$, which is almost equal to the projected distance of $30\,\mathrm{pc}$ for the Quintuplet cluster \citep{Figer1999}. Hence both clusters might have formed in the same environment, albeit at different times, and evolved in the strong tidal field of the Galactic centre. 
The slope of the mass function of the central part of the Arches cluster was first determined by \citet{Figer1999a} and found to be top-heavy with a slope of $\alpha = -1.65$ for $0.1 < R < 0.35\,\mathrm{pc}$. 
\citet{Stolte2002} found a slightly steeper slope of $\alpha = -1.8 \pm 0.2$ within $R < 0.4\,\mathrm{pc}$. Outside this radius the slope steepens to $\alpha = -2.70 \pm 0.7$, indicating again mass segregation towards the cluster centre. The authors correct for a radial extinction gradient outside of $0.2\,\mathrm{pc}$ and use the present-day masses determined from a $2\,\mathrm{Myr}$ Geneva isochrone.
A more recent study by \citet{Espinoza2009} finds a much steeper slope of $-2.1\pm0.2$ for $R < 0.4\,\mathrm{pc}$ consistent with a canonical IMF, but still a flattening towards the cluster core with  $\alpha = -1.88 \pm 0.20$ inside of $R = 0.2\,\mathrm{pc}$. \citet{Espinoza2009} account for differential extinction by individually dereddening the stars and infer initial masses instead of present-day masses from a $2.5\,\mathrm{Myr}$ Geneva isochrone.
\citet{Espinoza2009} have also shown that variations in the MF slope caused by the choice of metallicity and a wider range of cluster ages ($2.0-3.2\,\mathrm{Myr}$) are smaller than the fitting uncertainties. The steeper slopes are then most likely a consequence of individual dereddening each star prior to the stellar mass estimation. This suggests that individual dereddening is one of the most crucial aspects under variable extinction conditions to obtain realistic MF slopes. 
For the Quintuplet analysis presented above, individual dereddening was taken into account as well, and the initial stellar masses were used to create the MF. In this respect our slopes of the mass function of the Quintuplet cluster should be directly comparable with their results. 
However, as proper motions were not available, their membership selection is solely based on a strict colour-cut leaving the remaining contamination by field stars unclear.

For the Arches cluster, \citet{Kim2006} have quantified the effect of the internal cluster dynamics and the evaporation in the Galactic tidal field on the mass function measured within an annulus of $0.19-0.35\,\mathrm{pc}$.
Their Fokker-Planck calculations and N-body simulations yield a steepening of the mass function by $0.1$ to $0.2$ within the present cluster lifetime of $2.5\,\mathrm{Myr}$. 
At an older age of about $3-5\,\mathrm{Myr}$, the much more dispersed appearance of the Quintuplet with respect to the Arches cluster suggests that the Quintuplet cluster is dynamically more evolved and more affected by tidal effects.
While the dynamical evolution provides a tempting explanation for the flattened MF in the cluster centre, N-body simulations are required to confirm or disprove whether the flat  PDMF of the Quintuplet cluster can be explained by dynamical effects alone. 

\begin{table*}
\caption{Comparison of mass function slopes in the centres of Galactic young massive clusters.}
\label{t_slope_comparison}
\renewcommand\tabcolsep{4pt}
\begin{tabular}{lcccccccccc}
\hline\hline
Cluster name  & Age              & Distance         & Mass Range             & R               & $\alpha$       & R               & $\alpha$       & R               & $\alpha$       & References\\
	      & ($\mathrm{Myr}$) & ($\mathrm{kpc}$) & ($\mathrm{M_{\sun}}$)  & ($\mathrm{pc}$) &                & ($\mathrm{pc}$) &                & ($\mathrm{pc}$)                         \\
Quintuplet    & 3-5              &  8.0             &  $ > 5    $            &                 &                & $ < 0.5$        & $-1.66\pm0.15$ &                 &                & this work \\
Arches        & 2.5              &  8.0             &  $ > 10   $            & $ < 0.2$        & $-1.88\pm0.20$ & $ < 0.4$        & $-2.1\pm0.2$   &                 &                & 1 \\
NGC 3603      & 1-2.5            &  $6.0\pm0.8$     &  $ 4-20   $            & $ < 0.15$       & $-1.31$        & $ 0.3-0.4$      & $-1.72$        &                 &                & 2 \\
Westerlund 1  & 3-5              &  $3.55\pm0.17$   &  $ 3.4-27   $          &                 &                &                 &                & $ < 0.75$       & $-1.6$         & 3 \\        
\end{tabular}
\renewcommand\tabcolsep{6pt}
\tablebib{
(1) \protect{\citet{Espinoza2009}}; (2) \protect{\citet{Harayama2008}}; (3) \protect{\citet{Brandner2008}}
}
\end{table*}

\section{Summary and Outlook}
\label{s_summary_outlook}

We analysed high spatial resolution $H$- and $K_s$-band data of the Quintuplet cluster near to the Galactic centre obtained at the ESO/VLT with the NACO instrument. The cluster and the field star population were discerned  based on the individual proper motions determined from two $K_s$-band datasets with a time baseline of $5.0\,\mathrm{yr}$. Remaining contaminants were removed by a subsequent colour-cut, and spectroscopically identified M,K-giants from the Liermann spectral catalogue \citep{Liermann2009} were excluded. 
The slope of the present-day mass function of the Quintuplet cluster within $R<0.5\,\mathrm{pc}$ and $m_{\mathrm{init}} > 5\,\mathrm{M_{\sun}}$ was derived for the first time. The inital masses of the individually dereddened cluster members were determined from three Padova MS-isochrones of solar metallicity and ages of $3$, $4$ and $5\,\mathrm{Myr}$ as well as a $4\,\mathrm{Myr}$ Geneva isochrone to study the impact of using a different set of stellar evolution models. In order to avoid binning biases the mass function slopes were derived from bins with a uniform number of stars. This method produced robust results for each of the four isochrones. The derived slopes range from $\alpha = -1.52$ to $-1.74$, where the Salpeter slope is $-2.35$. The slope of the best fitting isochrone ($4\,\mathrm{Myr}$~ Padova) as well as the mean of all slopes for the four different isochrones are found to be $-1.66 \pm 0.14 $.

The orbital velocity and the internal velocity dispersion of the cluster were recently derived and constrain the formation and dynamical evolution in the GC tidal field (Stolte et al., in prep.). Using this velocity information, N-body simulations are currently undertaken (Harfst et al., in prep.) to probe if the disruptive effect of the GC tidal field can be claimed responsible for the shallow MF slope in the cluster core, or if a central overdensity of high mass stars had to be present initially during the formation of the cluster to explain the PDMF.
So far  the mass function of the Quintuplet cluster was only derived for its central part within $R<0.5\,\mathrm{pc}$. The question if the mass function steepens in the outer parts of the cluster as in the Arches or Westerlund~1, or if the global mass function remains top-heavy as in the case of NGC~3603, will be addressed in a future contribution. Due to the high field star density affecting the outer regions of the cluster, measured proper motions will be even more essential to derive a clean cluster sample than for the cluster core. These studies will contribute to solving the question whether star formation proceeds differently in the GC than in the spiral arms environment. 

\bibliographystyle{aa}		
\bibliography{hussmann_astroph_2011.bbl}

\clearpage

\end{document}